\documentclass[]{mn2e}
\usepackage{amssymb,psfig,graphicx}

\title[Systematic uncertainties in SED analysis]{Analysing observed star cluster SEDs with evolutionary synthesis models: Systematic uncertainties}

\author[P. Anders et al.]{P. Anders$^1$\thanks{E-mail:
panders@uni-sw.gwdg.de}, N. Bissantz$^2$, U. Fritze--v.  Alvensleben$^1$, R. de Grijs$^{3,4}$\\
$^1$ Universit\"ats-Sternwarte, University of G\"ottingen,
Geismarlandstr. 11, 37083 G\"ottingen, Germany, \\
$^2$ Institut f\"ur Mathematische Stochastik, University of G\"ottingen, Lotzestr. 13, 37083 G\"ottingen, Germany\\
$^3$ Institute of Astronomy, University of Cambridge, Madingley Road, Cambridge, CB3 0HA\\
$^4$ Department of Physics \& Astronomy, University of Sheffield, Hicks Building, Hounsfield Road, Sheffield, S3 7RH}

\date{Accepted ---. Received ---; in original form ---.}

\pubyear{2003}
\begin{document}
\maketitle

\begin{abstract}
We discuss the systematic uncertainties inherent to analyses of observed
(broad-band) {\bf S}pectral {\bf E}nergy {\bf D}istributions ({\bf SED}s) of
star clusters with evolutionary synthesis models. We investigate the effects
caused by restricting oneself to a limited number of available passbands,
choices of various passband combinations, finite observational errors, non-continuous
model input parameter values, and restrictions in parameter space allowed
during analysis. Starting from a complete set of {\sl UBVRIJH} passbands
(respectively their {\sl Hubble Space Telescope}/WFPC2 equivalents) we investigate
to which extent clusters with different combinations of age, metallicity,
internal extinction and mass can or cannot be disentangled in the various
evolutionary stages throughout their lifetimes and what are the most useful
passbands required to resolve the ambiguities. We find the {\sl U} and {\sl B}
bands to be of the highest significance, while the {\sl V} band and
near-infrared data provide additional constraints. A code is presented that
makes use of luminosities of a star cluster system in all of the possibly
available passbands, and tries to find ranges of allowed
age-metallicity-extinction-mass combinations for individual members of star
cluster systems. Numerous tests and examples are presented. We show the
importance of good photometric accuracies and of determining the cluster
parameters independently without any prior assumptions.
\end{abstract}

\begin{keywords}
globular clusters: general -- open clusters and associations: general --
galaxies: star clusters -- galaxies: evolution -- methods: data analysis
\end{keywords}

\section{Introduction}

Since the seminal work by Tinsley (1968), evolutionary synthesis has become a
powerful tool for the interpretation of integrated spectrophotometric
observations of galaxies and galactic subcomponents, such as star clusters.
Several groups introduced their evolutionary synthesis codes, e.g., Bruzual \&
Charlot (1993) [{\sc b\&c}], Leitherer et al. (1999) [{\sc starburst99}], Fioc
\& Rocca -- Volmerange (1997) [{\sc pegase}], Fritze -- v. Alvensleben \&
Gerhard (1994) [{\sc galev}] (all with regular updates), with various input
physics (evolutionary tracks vs. isochrones from various groups, different
sets of stellar spectral libraries, extinction laws ...). The codes do not
only vary in terms of input physics but also regarding computational
implementation, interpolation routines etc. A number of publications deals
with the intercomparison of various evolutionary synthesis codes (e.g. Worthey
1994, Charlot et al. 1996). The impact of uncertainties in the various model
parameters (such as in the descriptions of overshooting and mass loss, stellar
spectral libraries etc.) on the resulting colours is challenged by Yi (2003).
These publications find a good general agreement among the various models, and
assign acceptable uncertainties to the model results. Yi (2003) points out the
importance of a proper choice of filters for observing objects characterised
by different age ranges. This is justified by the light being dominated by
stars in different evolutionary stages at different times. The age-metallicity
degeneracy is a major drawback for accurate age determinations, especially for
young ages $\le$ 200 Myr.

{\bf In addition to the choice of the specific evolutionary synthesis model
used another important caveat merits discussion here.} A common assumption in
dealing with evolutionary synthesis is a well-populated stellar initial mass
function ({\bf IMF}), up to the model's upper mass limit. While this is
probably a justifiable assumption for galaxy-sized systems (although
uncertainties regarding the IMF {\sl slope} persist), it certainly breaks down
at levels of small (open) star clusters and OB associations, where stars are
formed purely stochastically (by consumption of the available amount of gas),
and these statistics dominate the observed dispersion in cluster luminosities.
A great deal of progress has been achieved already on this topic, in
particular by Cervi\~no and collaborators (e.g. Cervi\~no et al. 2002,
Cervi\~no \& Valls-Gabaud 2003). The main conclusion is that for systems more
massive than $\approx 10^5 M_\odot$ the impact of the stochasticity of the IMF
on the results is -- in general -- low, and the UV continuum is least affected
by stochastic dispersions.

{\bf The studies referred to before concentrated on the models themselves.
When comparing the model results with observations, in order to constrain the
cluster parameters age, metallicity, internal extinction, and mass, one does
not only need to take into account the model uncertainties, however. The final
parameter uncertainties also depend on the observational errors, the choice of
passbands used, their number, spectral coverage and individual filter
properties, and the analysis algorithm applied to one's data. The most common
way of model-observation comparison for astrophysical purposes is the
chi-squared minimisation technique, used e.g. for parameter determination of
star clusters (e.g. Maoz et al. 2001, de Grijs et al. 2003a,b), determination
of star formation histories of galaxies (e.g. Gavazzi et al. 2002), and
photometric redshift determination (e.g. Massarotti et al. 2001). Slightly
different, but comparable algorithms, like the least-squares method (e.g. Ma
et al. 2002) or maximum-likelihood estimation (e.g. Gil de Paz \& Madore 2002,
Bik et al. 2003), are used as well. However, see Bissantz \& Munk (2001) for a
critical discussion about the applicability of chi-squared versus
least-squares criteria.}

{\bf The aim of the present paper is a systematic evaluation of inherent
uncertainties in the analysis of observed star cluster spectral energy
distributions (SEDs) using evolutionary synthesis models. We define an SED as
an ensemble of (absolute) magnitudes in a given set of (broad-band) passbands.
We pay special attention to the most appropriate choice of passbands to
improve future observation strategies. We will point out severe pitfalls, such
as trends caused by finite observational errors and unjustified {\sl a priori}
assumptions.}

\section{Model description}
\label{sect.model}

{\bf In section \ref{sect1.model} we present the basic properties of our
evolutionary synthesis models. Section \ref{sect2.model} is a {\sl general}
description of our cluster SED analysis algorithm, regardless of whether it is 
used to study the parameters of observed star clusters or of simulated artificial 
clusters. In section \ref{sect3.model} we present the specific properties of the 
artificial clusters (clusters for which SEDs are taken directly from our models) 
used to simulate observed clusters and study the performance of our analysis tool.
From section \ref{sect.fit} onwards only these artificial clusters are used.}

\subsection{Input Models}
\label{sect1.model}

We use the single stellar population (SSP) models presented in Schulz et al.
(2002), with important improvements regarding the treatment of gaseous
emission in the early stages of the cluster evolution, as presented in Anders
\& Fritze -- v. Alvensleben (2003). These models include isochrones from the
Padova group including the TP-AGB phase, and model atmosphere spectra from
Lejeune et al. (1997; 1998). These extend from 90 {\AA} through 160 $\mu$m for
five different metallicities, $Z$ = 0.0004, 0.004, 0.008, 0.02 = $Z{_\odot}$
and 0.05 or [Fe/H] = $-$1.7, $-$0.7, $-$0.4, 0 and +0.4 (i.e., matching the
metallicities of the Padova isochrones), and gaseous emission (both lines and
continuum) due to the ionising flux from young massive stars. The models can
be retrieved from http://www.uni-sw.gwdg.de/$\sim$galev/panders/. For a
general description of the stellar models see Bertelli et al. (1994) and
Girardi et al. (2000), for details about the specific isochrones in our models
see Schulz et al. (2002). 

All calculations presented here are based on a Salpeter IMF in the mass range
of 0.15 to approximately 70 M$_\odot$ (0.15 to approx. 50 M$_\odot$ for
super-solar metallicity; following from the Padova isochrones). Stellar
synthesis models for a Scalo IMF are presented in Schulz et al. (2002) and
Anders \& Fritze -- v. Alvensleben (2003), and are available from the
aforementioned WWW address.

\subsection{General description of the analysis algorithm}
\label{sect2.model}

In order to analyse observed SEDs of star clusters in terms of the individual
cluster's age, metallicity, extinction, and mass we calculate a grid of models
for a large range of values for each of these parameters {\bf (except mass,
which is a simple scaling of the model mass [$M_{\rm model}$ = $1.6 \times
10^9 {\rm M}_\odot$] to the absolute observed cluster magnitudes)}. Input
parameters for the analysis are the time evolution of the spectra of the SSP
models, and the derived magnitude evolution in the various passbands.

{\bf The individual uncertainties contributing to the overall photometric
uncertainties are}: the observational uncertainties, an estimated model
uncertainty of 0.1 mag, and an uncertainty of an additional 0.1 mag for
passbands bluewards of the {\it B} band due to known calibration and model
problems in the UV. The total uncertainty is the square-root of the quadratic
sum of these individual errors. The observational and model uncertainties are
expected to be independent.

Galactic extinction is taken into account by dereddening the observations
using the Galactic extinction values from Schlegel et al. (1998).

First, we calculated dust-reddened spectra, using the starburst galaxy
extinction law by Calzetti et al. (2000), assuming a foreground screen
geometry,
\begin{center} {\small $k'(\lambda)=2.659 \times (-1.857 + 1.040 / \lambda) + 4.05$ \\ for 0.63 $\mu {\rm m} \le \lambda \le 2.20 \mu {\rm m}$, \\ ~ \\$k'(\lambda)=2.659 \times (-2.156+1.509/\lambda-0.198/\lambda^2+ 0.011/\lambda^3)+4.05$\\ for 0.09 $\mu {\rm m} \le \lambda < 0.63 \mu {\rm m}$} \end{center}
with a reddened flux 
\begin{center} $F_{{\rm red}}(\lambda)=F_0(\lambda) \times 10^{0.4  \times {\rm E}_s(B-V)   \times k'(\lambda)}$ \end{center}
and a range of values for the colour excess of the stellar continuum E$_s
(B-V)$. Since the gaseous emission is relevant only for a short time and even
then not the dominating term, the difference between the colour excess of the
stellar continuum and that from nebular gas emission lines (e.g. Calzetti et
al. 2000), is neglected.

We emphasise that the Calzetti law is valid only for starburst galaxies, while
for ``normal'' galaxies (i.e., undisturbed and quiescent spiral and elliptical
galaxies) it is probably at least marginally incorrect (due to the lower dust
content in such galaxies). However, for our systematic uncertainty analysis,
the specific shape of the extinction law assumed is of minor importance.

We construct SEDs from these models by folding the spectra with a large number
of filter response functions {\bf to obtain absolute magnitudes}. The
parameter resolutions are:
\begin{itemize}

\item Age: 4 Myr resolution for ages from 4 Myr -- 2.36 Gyr, 20 Myr resolution
for ages from 2.36 Gyr -- 14 Gyr;

\item Extinction: the resolution is $\Delta$E$(B-V) = 0.05$ mag, for E$(B-V) =
0.0 - 1.0$ mag;

\item Metallicities: [Fe/H] = $-1.7, -0.7, -0.4, 0$ and +0.4, as given by
the Padova isochrones;

{\bf \item Mass: an arbitrary model mass of $M_{\rm model} = 1.6 \times 10^9
{\rm M}_\odot$ is used.}
\end{itemize}

{\bf When comparing our observed SEDs with the model SEDs we first determine
the mass of the cluster by shifting the model SED onto the observed SED.}

\begin{figure*}
\begin{center}
\vspace{1.5cm}
\includegraphics[width=0.8\linewidth]{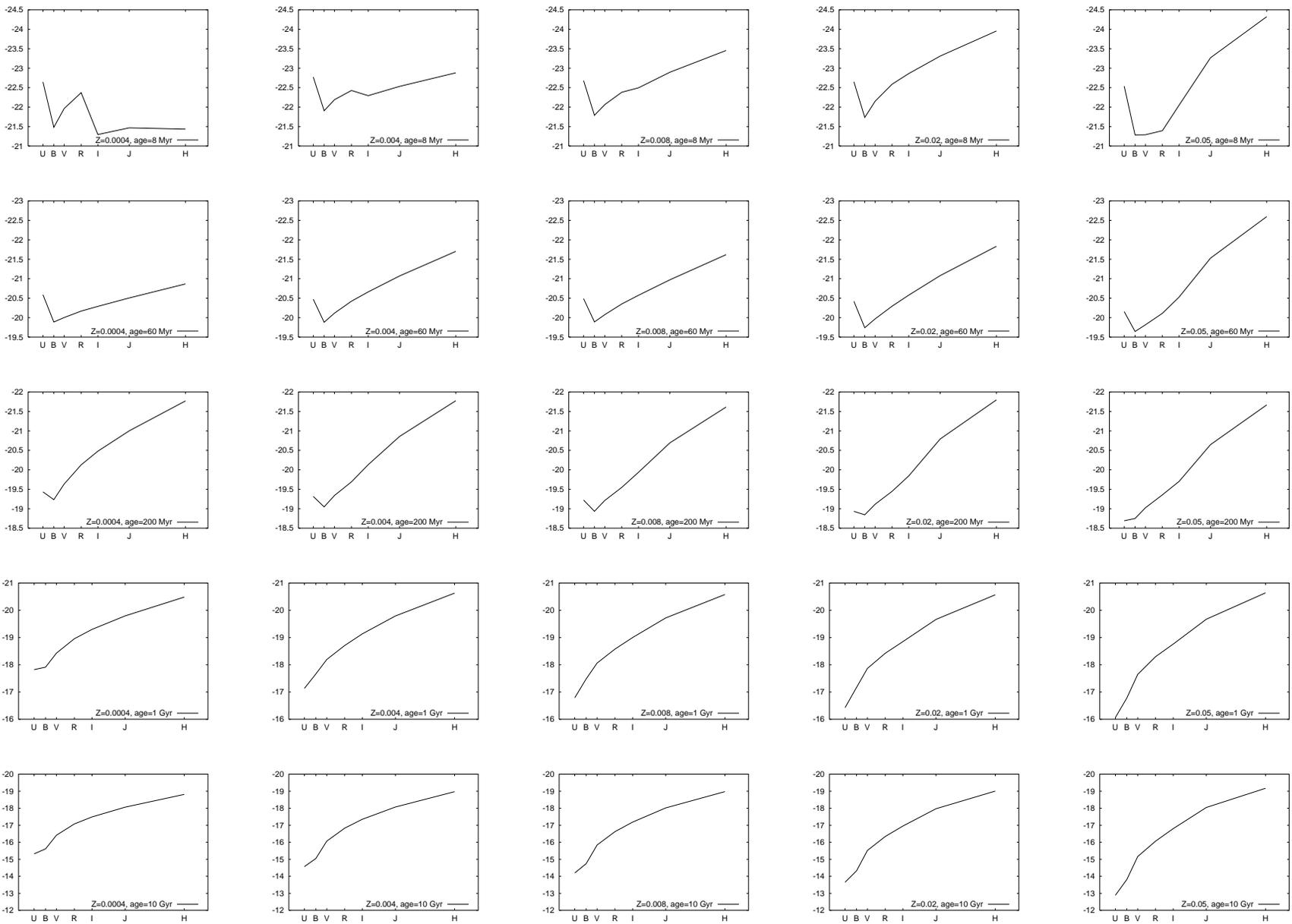}
\end{center}
\vspace{1cm}
\caption{\label{fig_SED}Representative SEDs, for the 5 available metallicities
and for 5 different, representative ages. The extinction is set to zero, and a
Salpeter IMF is used. {\bf We plot the absolute magnitudes in the respective
{\sl Hubble Space Telescope (HST)} passbands for $M_{\rm cluster} = M_{\rm
model}$ as a function of the effective wavelengths of the {\sl HST} passbands
(see section \ref{sect.pass}); the labels on the horizontal axis are the
corresponding standard Johnson passbands.}}
\end{figure*}

{\bf A number of these model SEDs (for $M_{\rm cluster} = M_{\rm model}$) are
shown in Fig. \ref{fig_SED}, for the 5 available metallicities and for 5
representative ages used for the artificial clusters considered in this paper
(see Section \ref{sect3.model}).}

Each of the models in our grid is now assigned a certain probability to be the
most appropriate one, given by a likelihood estimator of the form ${\rm p \sim
{\rm exp}(-\chi^2)}$, where $\chi^2=\sum{\frac{(m_{\rm obs}-m_{\rm
model})^2}{\sigma^2_{\rm obs}}}$, where $m_{\rm obs}$ and $m_{\rm model}$ are
the observed and the model magnitudes in each band, respectively, and
$\sigma_{\rm obs}$ are the observational uncertainties. The summation is over
all filters. Clusters with unusually large ``best'' $\chi^2$ are rejected,
since this is an indication of calibration errors, features not included in
the models (such as Wolf-Rayet star dominated spectra, objects younger than 4
Myr, etc.) or problems due to the limited resolution of the parameters. The
cut-off level is set to a total probability $\le 10^{-20}$, corresponding to a
${\rm \chi^2_{best} \ge 46}$. The total probability per cluster is then
normalised.

Subsequently, the model with the highest probability is chosen as the
``best-fit model''. Models with decreasing probabilities are summed up until
reaching 68.26 per cent total probability (= 1 $\sigma$ confidence interval)
to estimate the uncertainties in the best-fitting model. These uncertainties
are in fact upper limits, since their determination does not take into account
effects like the existence of several solution ``islands'' for one cluster
(such as e.g. the age-metallicity degeneracy, see below), and discretisation
in parameter space.

For real observations, several passband combinations (containing at least 4
passbands) were used for the analysis, to minimise the impact of calibration
errors and statistical effects. A minimum of 4 passbands is required to
determine the 4 free parameters age, metallicity, extinction and mass
independently (see also Anders et al. 2003, de Grijs et al. 2003a,b).

Only clusters with observational errors $\le$ 0.2 mag in all passbands of a
particular combination are included to minimise the uncertainties in the
results (except for some artificial clusters considered in this paper, for
which we adopt errors = 0.3 mag). For each combination, the best-fitting
models and their associated parameter uncertainties are determined. For a
given cluster all best-fitting models (and the associated uncertainties)
originating from the different passband combinations are compared. For each of
these best-fitting models the product P of the relative uncertainties

\begin{center}
P = ${\rm \frac{age^+}{age^-} \times \frac{mass^+}{mass^-} \times \frac{metallicity ~Z^+}{metallicity ~Z^-}}$
\end{center}

\noindent is calculated (the superscripts indicate the 1$\sigma$ upper ($^+$)
and lower ($^-$) limits, respectively). The relative uncertainty in the
extinction is not taken into account, since the lower extinction limit is
often zero. The data set with the lowest value of this product is adopted as
the most representative set of parameters (with its corresponding parameter
uncertainties) for the particular cluster being analysed. In cases where the
algorithm converges to a single model, a generic uncertainty of 30 per cent
for all parameters is assumed, in linear space, corresponding to an
uncertainty of $^{+0.1}_{-0.15}$ dex in logarithmic parameter space. See also
Anders et al. (2003) for an application to the star clusters in the dwarf
starburst galaxy NGC 1569, and de Grijs et al. (2003a,b) for applications of
this algorithm to clusters in the interacting starburst galaxies NGC 3310 and
NGC 6745.

\subsection{Artificial clusters}
\label{sect3.model}

In this study we will use artificial clusters to investigate the uncertainties
related to our analysis on the basis of a comparison with the model grid. {\bf
The SED magnitudes of the ``ideal'' artificial clusters are taken directly
from the models.} Standard parameters of these clusters are: metallicity
[Fe/H] = 0.0 = [Fe/H]$_\odot$, internal extinction E$(B-V) = 0.1$, and ages of
8 Myr (``cluster 1''), 60 Myr (``cluster 2''), 200 Myr (``cluster 3''), 1 Gyr
(``cluster 4''), and 10 Gyr (``cluster 5''). In this standard set only a age
variations, and neither metallicity nor extinction variations are considered
initially, for reasons of clarity. The impact of varying the metallicity and
extinction values is treated separately, see especially Sect.
\ref{sect.varinput}. The cluster mass is the model's mass $1.6 \times 10^9
{\rm M}_\odot$, the ``observational'' errors are set to be 0.1 mag in each
filter. Unless otherwise indicated, the clusters in this paper will have these
standard parameters.

For each of these 5 sets of artificial cluster parameters 10,000 cluster SEDs
were generated by adding statistical noise to the magnitudes of the ``ideal''
cluster. The errors are drawn from a Gaussian distribution with the Gaussian
$\sigma$ corresponding to the ``observational'' uncertainty (= 0.1 mag as
standard value).

All clusters are analysed separately with our algorithm in order to assess
under which conditions and to what accuracy their input parameters are
recovered by our method. Subsequently, all clusters originating from a given
``ideal'' cluster are used to calculate median parameters and their associated
uncertainties. The uncertainties are centred around the median solution; they
serve as equivalents to the 1$\sigma$ standard deviation around the average
values. However, for our analysis we chose to use the median instead of the
average of the distribution, since we believe the median to be physically more
relevant. We are interested in finding the most likely result when comparing
our model grid with observations.

Free parameters are the metallicity [Fe/H], the extinction E$(B-V)$, log(age)
and log(mass). [Fe/H] and log(age) are used instead of Z and age because the
evolution of magnitudes is approximately linear in [Fe/H] and log(age).


\section{Study of the accuracy of our analysis}
\label{sect.fit}

\subsection{Passbands included in our analysis}
\label{sect.pass}

We consider the following filters (the impact of only slightly different
filter response curves is small). All filters are taken from the set of
available filters for observations of the {\sl Hubble Space Telescope
(HST)}/WFPC2, ACS, and NICMOS cameras.

The standard set of filters is: {\sl HST} WFPC2 (and ACS) filters F336W
(``{\sl U}''), F439W (``{\sl B}''), F555W (``{\sl V}''), F675W (``{\sl R}''),
F814W (``{\sl I}''), NICMOS (NIC2 camera) F110W (``{\sl J}''), F160W (``{\sl
H}''). This standard set will be referred to as ``{\sl UBVRIJH}''. In addition
the following filters are included in our study as well: the {\sl HST} WFPC2
(and ACS where appropriate) wide filters F300W (``wide {\sl U}''), F450W
(``wide {\sl B}''), F606W (``wide {\sl V}''), F702W (``wide {\sl R}'') and the
{\sl HST} Str\"omgren filters F336W (``{\sl u}'' $\equiv$ ``{\sl U}''), F410M
(``{\sl v}''), F467M (``{\sl b}''), F547M (``{\sl y}'').

{\bf In this paper we will use the term ``UV passband'' essentially for the U
band, and the term ``NIR passbands'' for the J and H bands.}

In the relevant figures, the horizontal lines mark the input values, and the
symbols represent the median of the recovered values with the associated
uncertainties. The clusters with ``cluster number'' = 1 $\le$ x $<$ 2 are
clusters with the youngest input age of 8 Myr, clusters with ``cluster
number'' = 2 $\le$ x $<$ 3 are clusters with an input age of 60 Myr, and so on
(this offset is chosen for reasons of clarity).

\subsection{Choice of passband combination}

First, we investigate which passbands contain the maximum amount of
information, and hence which passbands are preferred for observations, if one
can obtain observations in only a limited number of passbands. This aims at
improving future observing strategies.

\begin{figure*}
\begin{center}
\includegraphics[angle=-90,width=0.8\linewidth]{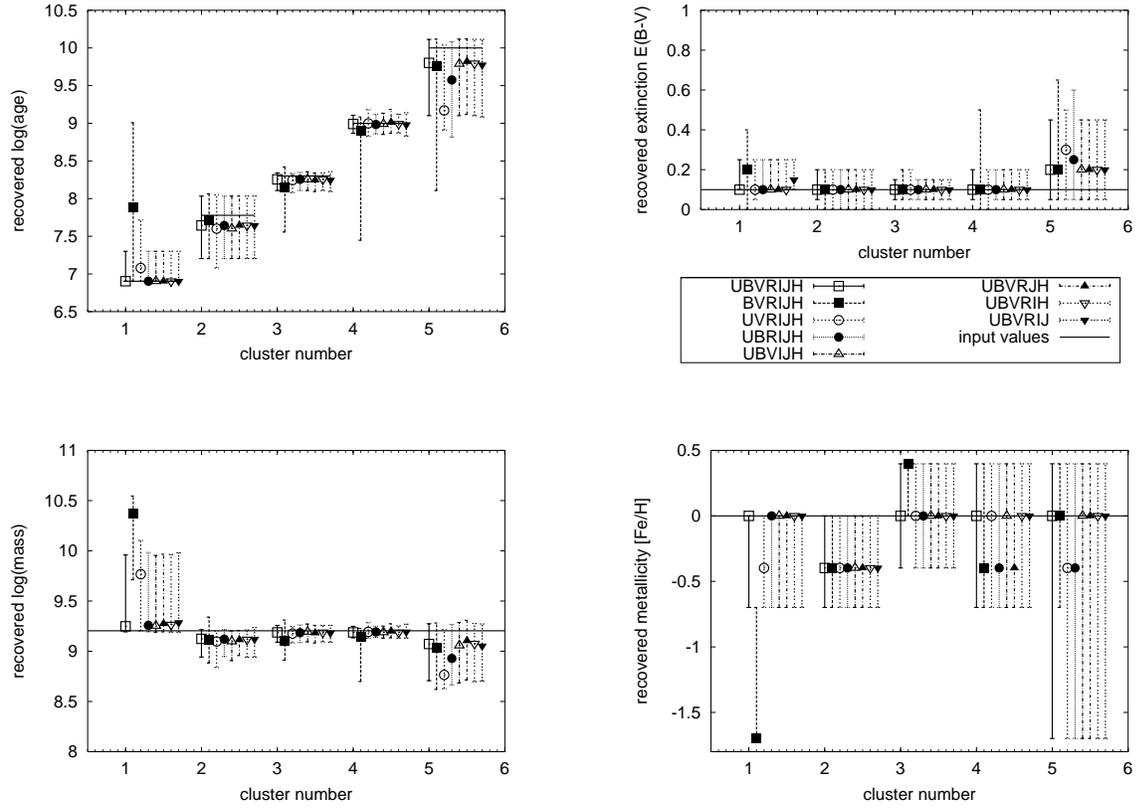}
\end{center}
\vspace{0.3cm}
\caption{Dispersion of recovered properties of artificial clusters, assuming availability of {\sl UBVRIJH} and passband combinations rejecting one of the {\sl UBVRIJH} passbands, as indicated in the legend. Cluster parameters are standard.}
\label{fig_76}
\end{figure*}

\subsubsection{Importance of individual passbands}
\label{sect.UBVRIJH}

In Fig. \ref{fig_76} we present the dispersions in our recovered parameters
using the standard input parameters, and SEDs covering the full wavelength
range {\sl UBVRIJH}, compared with passband combinations where one of the {\sl
UBVRIJH} passbands is left out.

This figure provides direct evidence of the importance of the {\sl U} band
(and to a lesser degree also of the {\sl B} band) for all stages of cluster
evolution, while for ages $>$ 1 Gyr also a lack of the {\sl V} band results in
problems to recover the age. The systematic deviations from the input values
for the combinations without the {\sl U} or {\sl B} bands are caused by an
insufficiently accurate determination of the cluster metallicity. The
resulting SED changes are therefore balanced by the analysis algorithm by
adjusting the extinction and/or age, and are also accompanied by
higher-than-input median masses in our fit results.

Systematic biases are only apparent in the age determination of the oldest
artificial cluster (with a slight bias towards younger recovered ages),
balanced by an overestimate of the internal extinction (which is a sign of the
age-extinction degeneracy) and a minor bias towards smaller median masses. For
the 60 Myr-old artificial cluster, the metallicity determination leads to an
underestimate (presumably due to the criss-crossing of the models and/or the
non-negligible impact of the age-metallicity degeneracy at these ages) for
all passband combinations, while for the oldest cluster the uncertainty in the
metallicity determination encompasses almost the entire available range.

In general, the median values recovered by our code agree fairly well with the
input parameters, with the exceptions mentioned above. The parameter
dispersions are largest for the young (ages $\le$ 60 Myr) and the oldest (age
= 10 Gyr) clusters. This is caused by the criss-crossing of the models for
young ages and the flat magnitude evolution for old ages.

The importance of the {\sl U} and {\sl B} band is immediately apparent from
the overview of artificial SEDs presented in Fig. \ref{fig_SED}. {\sl U} and
{\sl B} are important for tracing the hook-like structure for young ages,
while there appears to be a kink in the SEDs at the {\sl V} band for older
ages.

\subsubsection{Combinations of 4 passbands}
\label{sect.4pass}

The minimum number of passbands required to determine the 4 free parameters
age, metallicity, extinction and mass independently is four.

\begin{figure*}
\begin{center}
\includegraphics[angle=-90,width=0.8\linewidth]{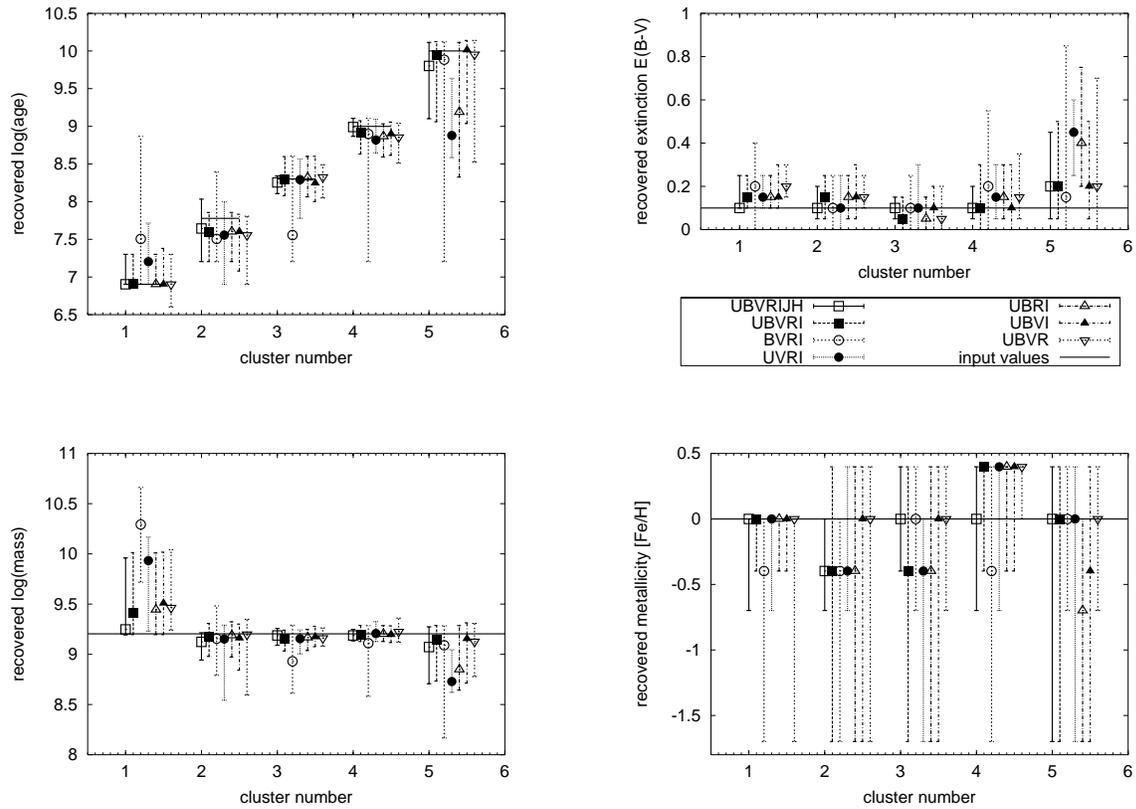}
\end{center}
\vspace{0.3cm}
\caption{Dispersion of recovered properties of artificial clusters, assuming availability of various optical passband combinations, as indicated in the legend. Cluster parameters are standard.}
\label{fig_74I}
\end{figure*}

\begin{figure*}
\begin{center}
\includegraphics[angle=-90,width=0.8\linewidth]{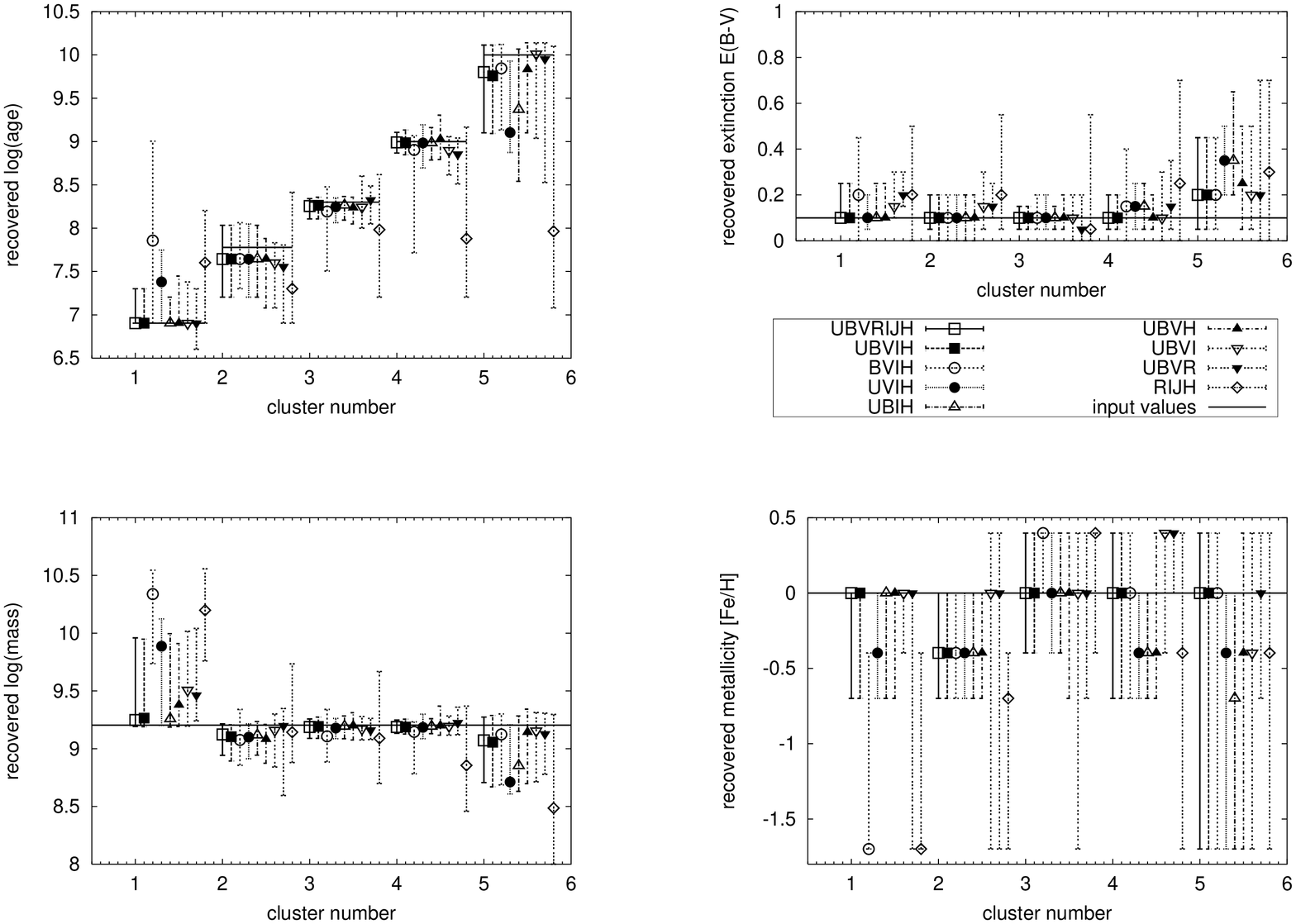}
\end{center}
\vspace{0.3cm}
\caption{Dispersion of recovered properties of artificial clusters, assuming availability of various optical+NIR passband combinations, as indicated in the legend. Cluster parameters are standard.}
\label{fig_74H}
\end{figure*}

In Figs. \ref{fig_74I} and \ref{fig_74H} we present the recovered
parameters for {\sl UBVRIJH} compared with various passband combinations
consisting of 4 passbands, for optical filters only and including one
near-infrared (NIR) band, respectively.

For optical passbands only, the {\sl U} band plays a major role once more,
especially in determining the metallicity. Missing {\sl U}-band information
leads to underestimates of the metallicity, thereby causing extinction values
and ages to be adjusted improperly, and hence this also leads to incorrect
mass estimates. Even in cases where the median is recovered correctly, such
clusters show the largest uncertainties. In some cases, missing {\sl B}-band
information has similar effects, especially for the youngest cluster, while
for the oldest cluster the {\it B} band is vital to break the age-extinction
degeneracy. Only for the oldest cluster, the {\sl V} band contains vital
information, which is in accordance with our results in Sect.
\ref{sect.UBVRIJH}.

For optical+NIR passbands, the situation is similar: The {\sl U} band (and to
a lesser degree also the {\sl B} band) is essential. Generally, the offsets
from the input values and the uncertainty ranges are smaller than for optical
passbands only, thus proving the importance of NIR data. Choosing a NIR band
closely resembling the {\sl K} band instead of {\sl J} or {\sl H} would give
similar results, possibly restricting the values slightly better. However, we
concentrated on the {\sl H} band since there are more observations available
in {\sl H} in the {\sl HST} data archive than for filters with longer
central wavelengths.

{\bf In Fig. \ref{fig_74H} we also see the effect of a limited wavelength
coverage: in all parameters, the {\sl RIJH} combination gives the worst
results (see also de Grijs et al. 2003a). Similar, but less pronounced is the
effect for the {\sl UBVR} combination.}

\begin{figure*}
\begin{center}
\includegraphics[angle=-90,width=0.8\linewidth]{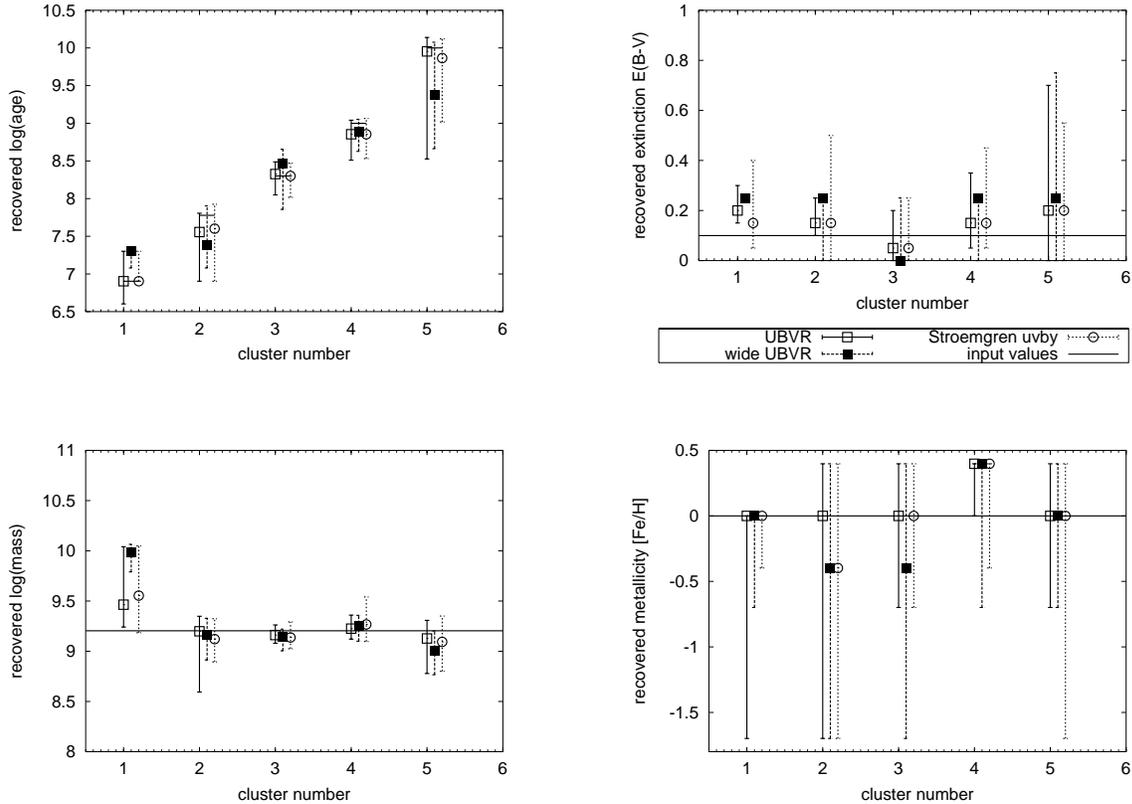}
\end{center}
\vspace{0.3cm}
\caption{Dispersion of recovered properties of artificial clusters, comparing various wide and medium-band {\sl HST} filters, as indicated in the legend. Cluster parameters are standard.}
\label{fig_misc}
\end{figure*}

Fig. \ref{fig_misc} compares the normal WFPC2 {\sl UBVR} system with the
corresponding passband combination using the WFPC2 wide filters. In addition,
results based on the medium band Str\"omgren filter system of WFPC2 are shown.

In most cases the wide filter system gives slightly worse results than the
standard system. However, driven by the wider filter response curves and the
associated smaller observational errors thanks to the larger flux throughput,
the wide system might be preferable, e.g., for faint objects.

Using the WFPC2 Str\"omgren medium-band system does not result in significant
improvements compared to wide-band systems. In conjunction with the lower flux
throughput (caused by the narrower bandwidth) this system seems less
preferable for our purpose. We emphasise that this only holds for our SED
analysis.

In de Grijs et al. (2003a) we investigated the impact of the choice of
passbands for the young cluster system (with ages of few $\times 10-100$ Myr)
in NGC 3310 with {\sl HST} data from the UV through to the NIR. Starting with
the full set of available passbands, we studied the changes in accuracy of the
recovered parameters if we repeated the analysis using only a subset of our
passbands. By comparing the results from our analyses using all passbands with
those from smaller subsets we found severe biases in the age distributions
originating from different passband combinations, in particular for
combinations biased towards longer wavelengths ({\sl VIJH}), but also for {\sl
UV-UBV} (covering shorter wavelengths only) and {\sl BVIJH}, consistent with
the results presented here.

\subsubsection{Conclusions on the choice of passbands}

{\bf From these comparisons we conclude that the passband combinations for the
most reliable parameter determination must include the {\sl U} band, the {\sl
B} band, and use the maximum available wavelength range, preferably including
at least one NIR band.} If only observations in 4 passbands can be obtained,
the best combinations are {\sl UBIH} or {\sl UBVH}, especially for genuinely
old objects, and {\sl UBVI}, if NIR data cannot be acquired. We emphasise once
again that tracing the kink around the {\sl B} /{\sl V} band in the SEDs (see
Fig. \ref{fig_SED}) is vital. For improved metallicity determinations, and
consequently for improved determinations of the other parameters as well, NIR
data seem to be crucial (for young clusters the {\sl U} /{\sl B} bands are
also important, in order to determine the metallicity correctly). However, due
to the limited metallicity resolution (and the numerous effects the
metallicity has on the synthetic magnitudes), the metallicity determination
remains the weakest point in our cluster analysis algorithm, and presumably in
any routine using synthetic magnitudes from stellar isochrones or tracks.

\subsection{Varying the input parameters}
\label{sect.varinput}

In this section we investigate to which extent the input parameters can be
recovered as a function of their respective values and observational errors.

\subsubsection{Using all 7 filters}

\begin{figure*}
\begin{center}
\includegraphics[angle=-90,width=0.8\linewidth]{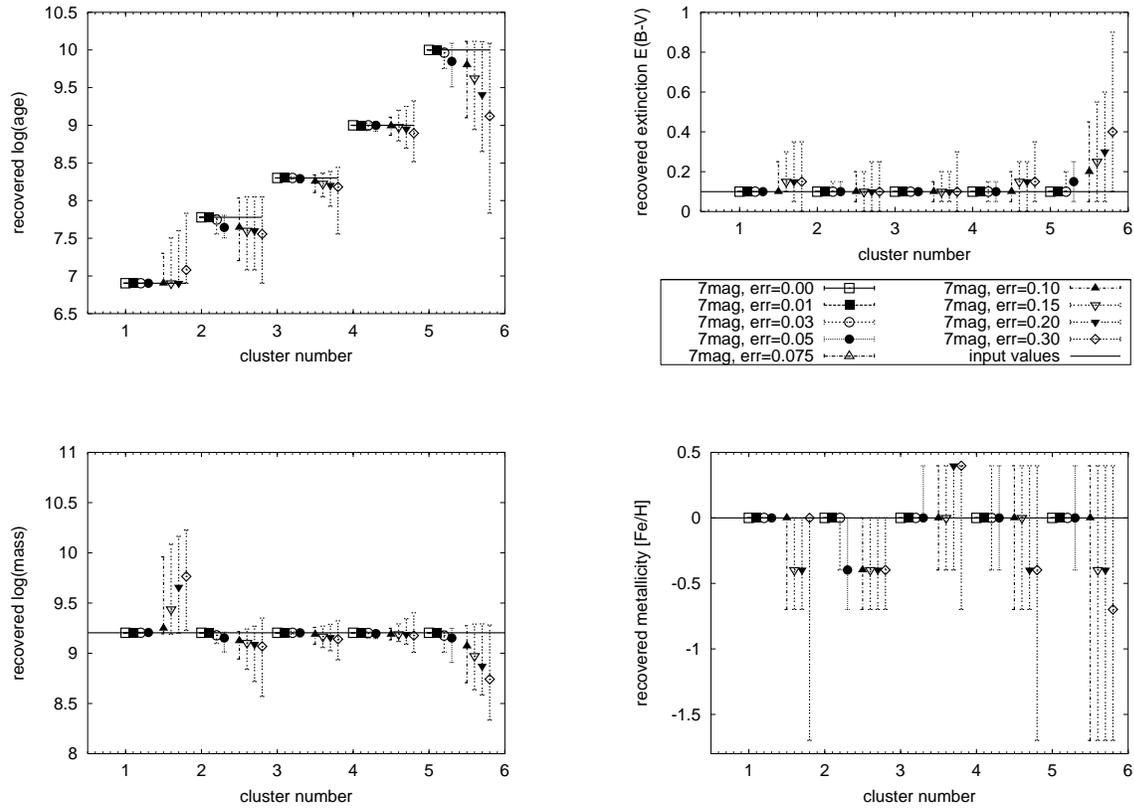}
\end{center}
\vspace{0.3cm}
\caption{Dispersion of recovered properties of artificial clusters, assuming availability of {\sl UBVRIJH} magnitudes and varying observational errors, as indicated in the legend. Other parameters are standard.}
\label{fig_7err}
\end{figure*}

Fig. \ref{fig_7err} shows, for a range of observational uncertainties, the
reliability of our recovered parameters if the standard set of filters ({\sl
UBVRIJH}) is available. We caution that we still apply the model uncertainty
of 0.1 mag (and an additional uncertainty of 0.1 mag for UV passbands).

A slight trend towards an underestimate of the ages, balanced by a slight
overestimate of the internal extinction and an occasional underestimate of the
metallicity, is seen. However, even for the largest observational errors of
0.3 mag that we tested for, all recovered parameters are consistent with the
input parameters, within the uncertainties.

With increasing observational errors, there seems to be a trend to
underestimate the ages for the oldest cluster, balanced by an increasing
overestimate of the internal extinction. For genuinely old cluster systems,
this degeneracy can be broken by restricting the extinction range. This is
generally justified, since such systems are usually dust-poor, if not
dust-free, and show fairly homogeneous extinction distributions.

\begin{figure*}
\begin{center}
\includegraphics[angle=-90,width=0.8\linewidth]{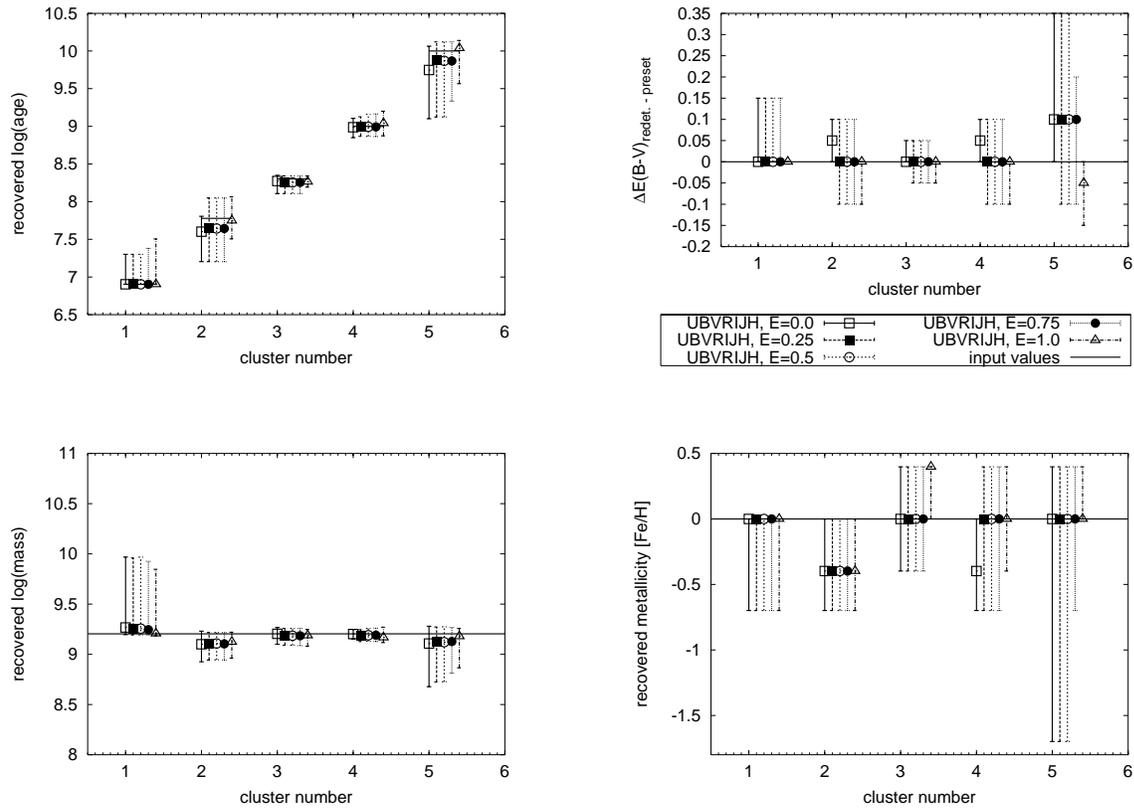}
\end{center}
\vspace{0.3cm}
\caption{Dispersion of recovered properties of artificial clusters, assuming availability of {\sl UBVRIJH} magnitudes and varying internal extinction values, as indicated in the legend. Other parameters are standard.}
\label{fig_7ext}
\end{figure*}

Fig. \ref{fig_7ext} shows that the degree to which our code recovers the input
parameters is largely independent of the input extinction value, with the
exception of the ages recovered for the oldest artificial clusters (in this
latter case clear signs of the age-extinction degeneracy are apparent). The
remaining deviations of the median recovered values from the input values are
always less than 0.2 dex, and in most cases even smaller. The deviations in
metallicity and extinction are one step in resolution (except for the
extinction of the oldest cluster, which is, in most cases, 2 steps off). Small
trends for increasing age underestimates with lower input extinction are
discernible.

\begin{figure*}
\begin{center}
\includegraphics[angle=-90,width=0.8\linewidth]{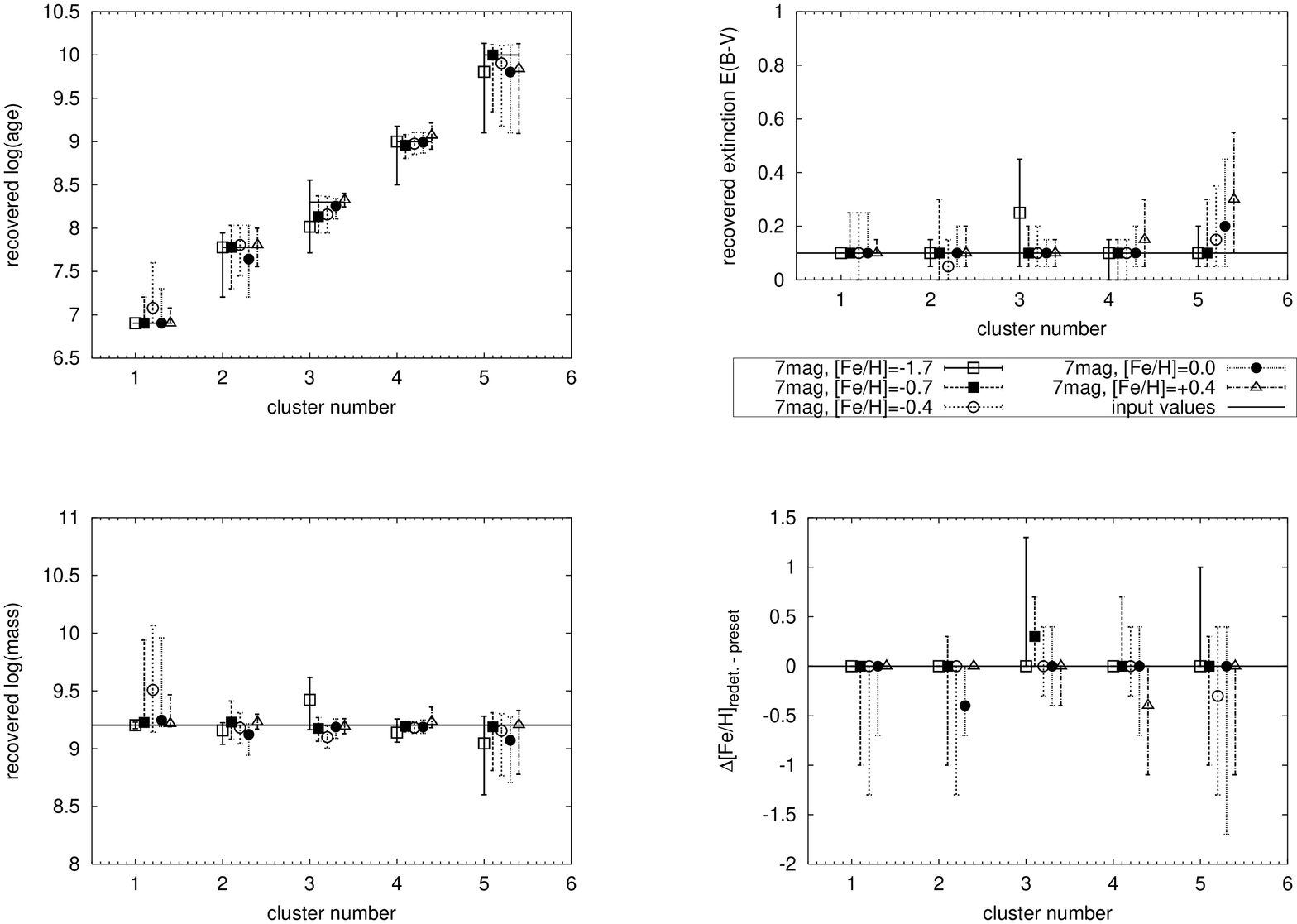}
\end{center}
\vspace{0.3cm}
\caption{Dispersion of recovered properties of artificial clusters, assuming availability of {\sl UBVRIJH} magnitudes and varying metallicity values, as indicated in the legend. Other parameters are standard.}
\label{fig_7met}
\end{figure*}

Fig. \ref{fig_7met} indicates good agreement between the input parameters and
their recovered values for all 5 metallicities. Median extinction values and
metallicities match the input values very well. The age determination is
correct to $\Delta$ log(age) $\la$ 0.25 dex. The mass is recovered very well,
as is the extinction. The various metallicity input values are in general
correctly recovered, only in few cases a difference of one resolution step is
seen.

\subsubsection{Using the minimum of 4 filters}

The following figures show the accuracy if observations in only the minimum
of 4 passbands are available (i.e., a more realistic case). We discuss the
best-suited 4-passband-combination identified in Sect. \ref{sect.4pass},
including the {\sl H} band, i.e. the combination {\sl UBIH}.

\begin{figure*}
\begin{center}
\includegraphics[angle=-90,width=0.8\linewidth]{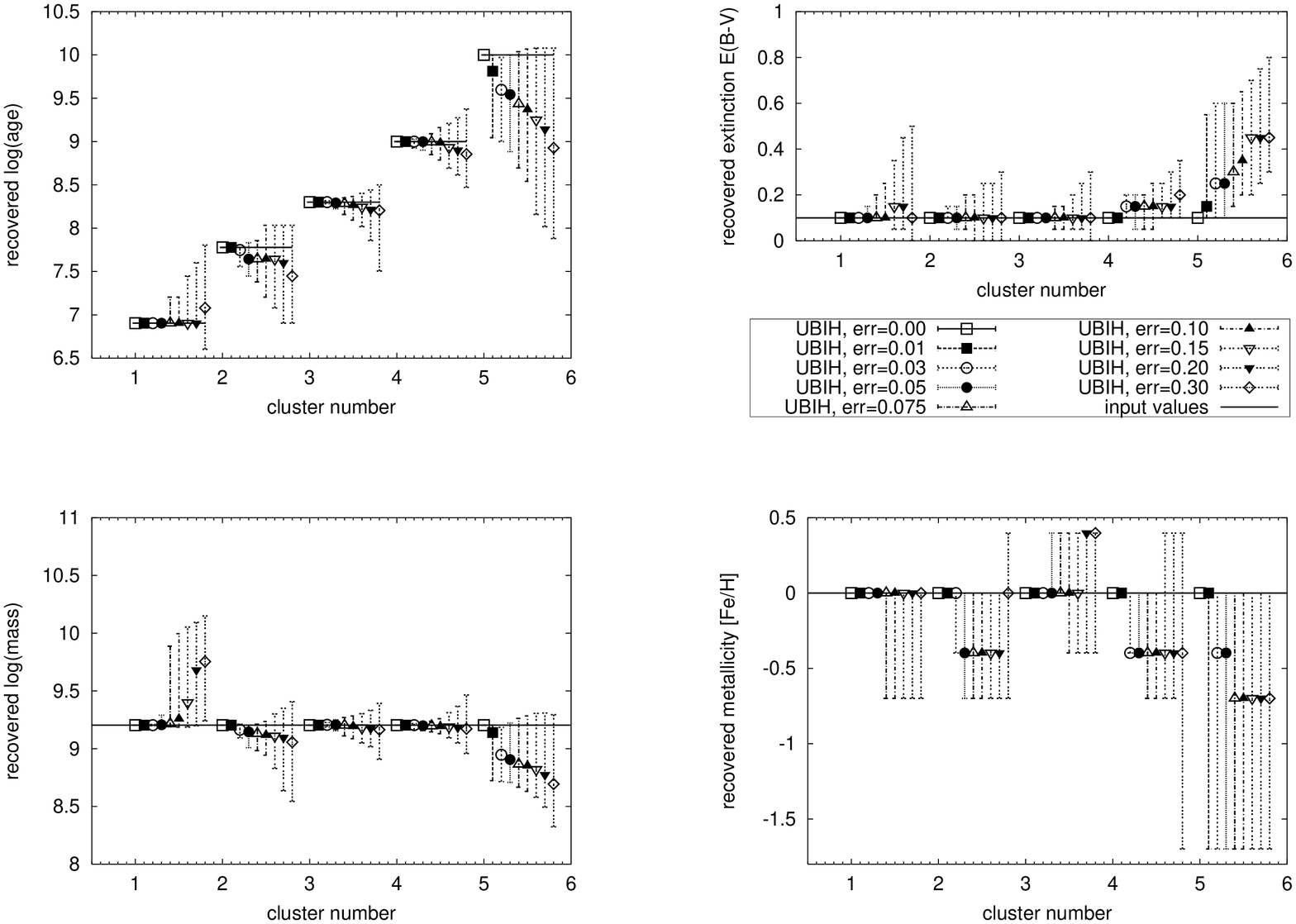}
\end{center}
\vspace{0.3cm}
\caption{Dispersion of recovered properties of artificial clusters, assuming availability of {\sl UBIH} magnitudes and varying observational errors, as indicated in the legend. Other parameters are standard.}
\label{fig_4err}
\end{figure*}

Fig. \ref{fig_4err} shows significant trends caused by increasing
observational errors, especially for the oldest clusters. For the other
clusters, the trends are less severe, with deviations of less than a factor of
2, or one step in the metallicity resolution. For the oldest cluster, the
deviations are up to 1 dex in age, 2 steps in metallicity, 0.35 mag in
E$(B-V)$ and 0.5 dex in mass, for the largest observational errors, i.e. 0.3
mag.

\begin{figure*}
\begin{center}
\includegraphics[angle=-90,width=0.8\linewidth]{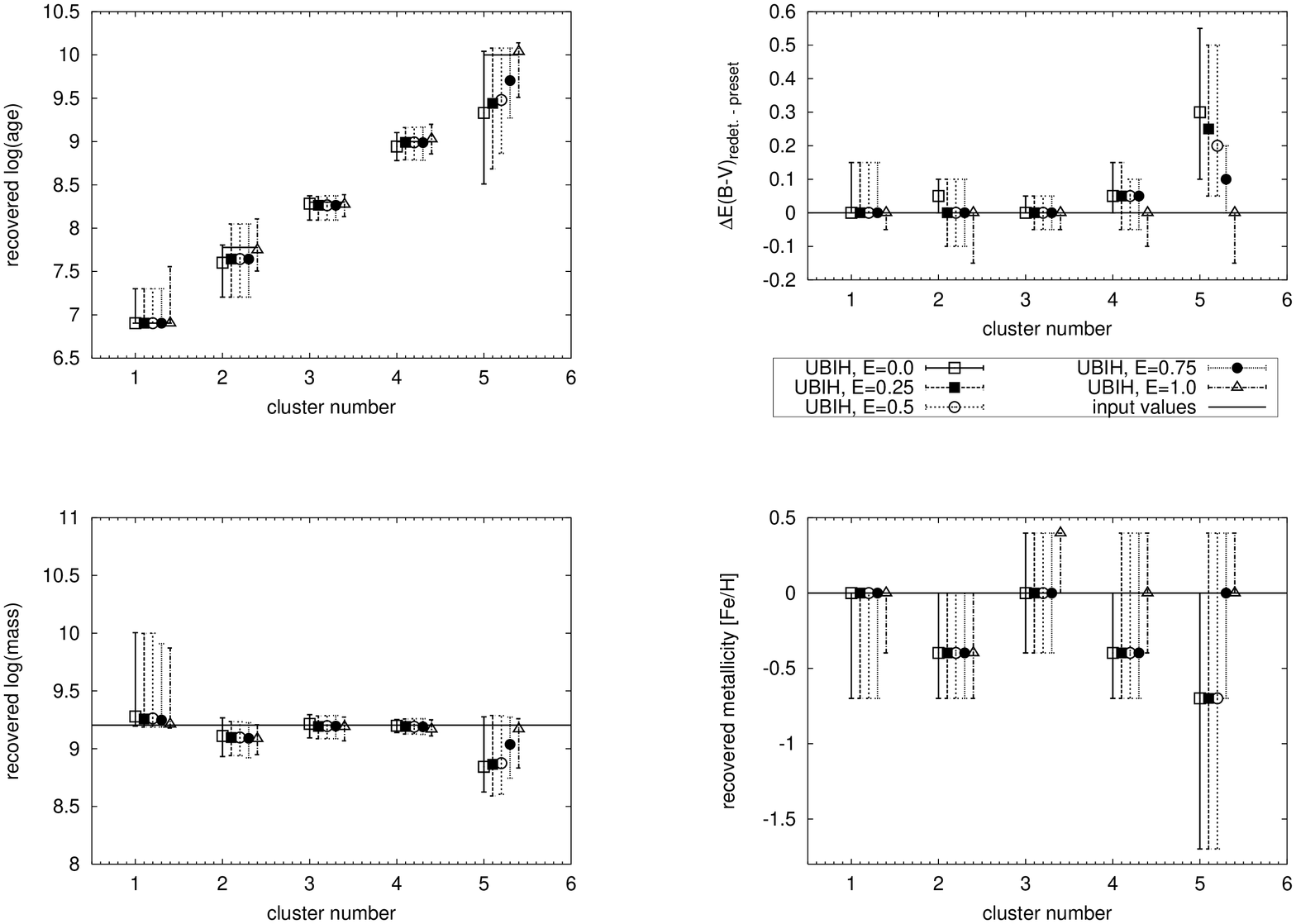}
\end{center}
\vspace{0.3cm}
\caption{Dispersion of recovered properties of artificial clusters, assuming availability of {\sl UBIH} magnitudes and varying internal extinction values, as indicated in the legend. Other parameters are standard.}
\label{fig_4ext}
\end{figure*}

Fig. \ref{fig_4ext} shows the recovered values for {\sl UBIH} and various
input extinction values. For all but the oldest cluster, the recovered
parameters reproduce the input values very well. The offsets and uncertainties
are slightly larger, but comparable to the corresponding values for {\sl
UBVRIJH}.

For the oldest cluster there are pronounced trends: with increasing input
extinction, the recovered age, metallicity and mass estimates increase, while
the offsets of the recovered extinction values from their input values
decrease. We find that the higher the input extinction, the better all input
parameters are recovered.

\begin{figure*}
\begin{center}
\includegraphics[angle=-90,width=0.8\linewidth]{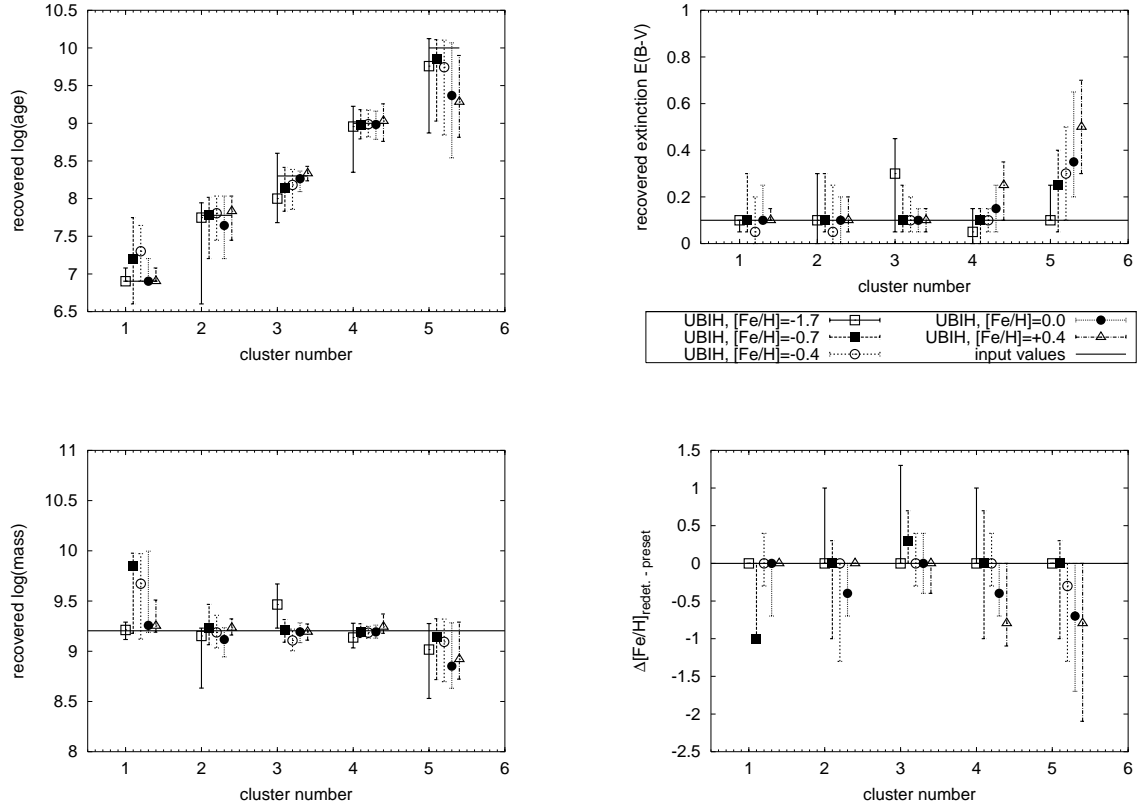}
\end{center}
\vspace{0.3cm}
\caption{Dispersion of recovered properties of artificial clusters, assuming availability of {\sl UBIH} magnitudes and varying metallicity values, as indicated in the legend. Other parameters are standard.}
\label{fig_4met}
\end{figure*}

Fig. \ref{fig_4met} shows the recovered values for {\sl UBIH} and various
input metallicities. The trends with increasing input metallicity are less
obvious, except again for the oldest cluster (more significant metallicity
underestimates and extinction overestimates with increasing input
metallicity). This behaviour is also present, but less pronounced for the
second oldest (= 1 Gyr old) cluster. For the other, younger clusters, the
behaviour appears almost random. This is caused by the slow and steady
evolution of magnitudes at large ages: with increasing metallicity the
magnitudes become fainter and the colours redder. For younger ages the
evolution of the magnitudes is less linear, and they criss-cross several
times.

\subsubsection{Conclusions on the impact of varying the input parameters}

We have investigated the impact of varying the input parameters on the
accuracy of our parameter recovery. We find very good agreement, with
generally small deviations, by either varying the input extinction or the
input metallicity. Only for the oldest artificial clusters there are
significant trends in the recovered values with increasing input extinction
and metallicity. This can be attributed to a number of degeneracies. {\bf We
remind the reader that the upper age limit of the evolutionary synthesis
models is 14 Gyr.}

For increasing observational uncertainties there are clear trends of
increasing recovered extinction and decreasing ages (with a small increase for
the youngest cluster only), mass (an increase for the youngest cluster only)
and metallicity. The uncertainties increase as well, as expected.

The results using either {\sl UBVRIJH} or {\sl UBIH} are fairly similar. Using
4 passbands only slightly increases the offsets of the median recovered values
from the input values, as well as the uncertainties. Some trends, especially
for the oldest cluster, become more significant.


\subsection{Restricting the parameter space to the correct ranges}

In this section we investigate the consequences of {\sl a priori} restrictions
of the parameter space. This might make sense in cases where, e.g., large
observational errors may inhibit reasonable parameter constraints or where
additional information is available, such as spectroscopic abundances,
dynamical age estimates for the starburst event that induced the cluster
formation, a low-metallicity dwarf galaxy environment, etc. Here, we explore
various cases where we restrict some of our free parameters to the (correct)
range of input values, and use passband combinations {\sl UBVRIJH} and
combinations lacking one of the {\sl UBVRIJH} passbands to recover the input
parameters.

\begin{figure*}
\begin{center}
\includegraphics[angle=-90,width=0.8\linewidth]{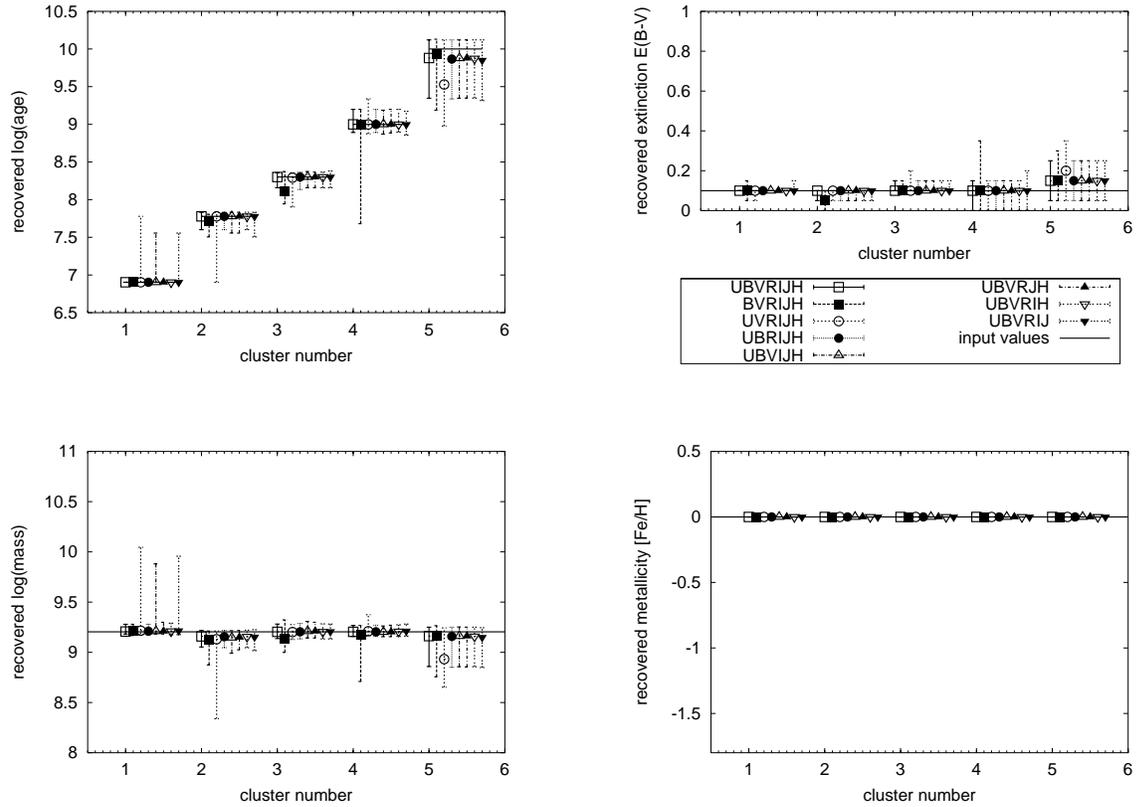}
\end{center}
\vspace{0.1cm}
\caption{Dispersion of recovered properties of artificial clusters, assuming availability of {\sl UBVRIJH} and passband combinations without one of the {\sl UBVRIJH} passbands, as indicated in the legend. Solutions were sought with metallicity fixed to the input value. Cluster parameters are standard.}
\label{fig_76Z}
\end{figure*}

\begin{figure*}
\begin{center}
\includegraphics[angle=-90,width=0.8\linewidth]{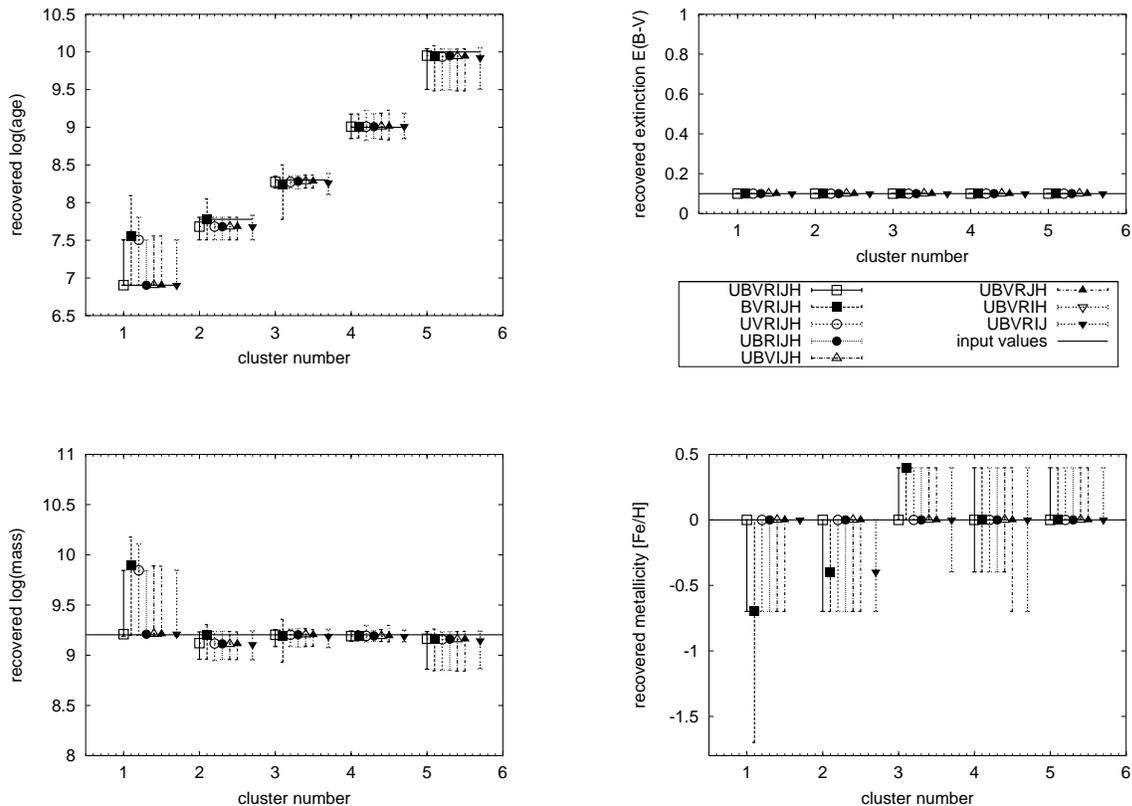}
\end{center}
\vspace{0.1cm}
\caption{Dispersion of recovered properties of artificial clusters, assuming availability of {\sl UBVRIJH} and passband combinations without one of the {\sl UBVRIJH} passbands, as indicated in the legend. Solutions were sought with extinction fixed to the input value. Cluster parameters are standard.}
\label{fig_76E}
\end{figure*}

Fig. \ref{fig_76Z} shows the results when we restrict the metallicity to the
(input) solar metallicity. Apart from the oldest cluster (which shows a slight
underestimate of the age, balanced by a slight overestimate of the extinction)
the recovered values agree almost perfectly with the input values, in any case
much better than without metallicity restriction (cf. Fig. \ref{fig_76}). The
importance of the {\sl U} and {\sl B} bands, and of the largest possible
wavelength coverage are still reflected by the larger uncertainties for
observations lacking those filters. The deviations for the oldest cluster are
a result of the age-extinction degeneracy.

In Fig. \ref{fig_76E} we investigate the consequences of restricting the
analysis to the input extinction, allowing variations only in metallicity and
age. The deviations of the median solutions from the input values, and the
uncertainty ranges are significantly reduced compared to the unrestricted case
shown in Fig. \ref{fig_76}. Especially for the oldest cluster, some
degeneracies are removed and the recovered values agree much better with the
input values than in the unrestricted case. For genuinely old cluster systems
the assumption of a generic low extinction may be justified, since such
systems are common in old relaxed galaxies with generally low (and uniform)
dust content.

Nevertheless, the importance of including the {\sl U} and {\sl B} bands is
still apparent, especially in the age determination for the youngest cluster.
By comparison with Fig. \ref{fig_76Z} we can attribute this behaviour to the
age-metallicity degeneracy.

\begin{figure*}
\begin{center}
\includegraphics[angle=-90,width=0.8\linewidth]{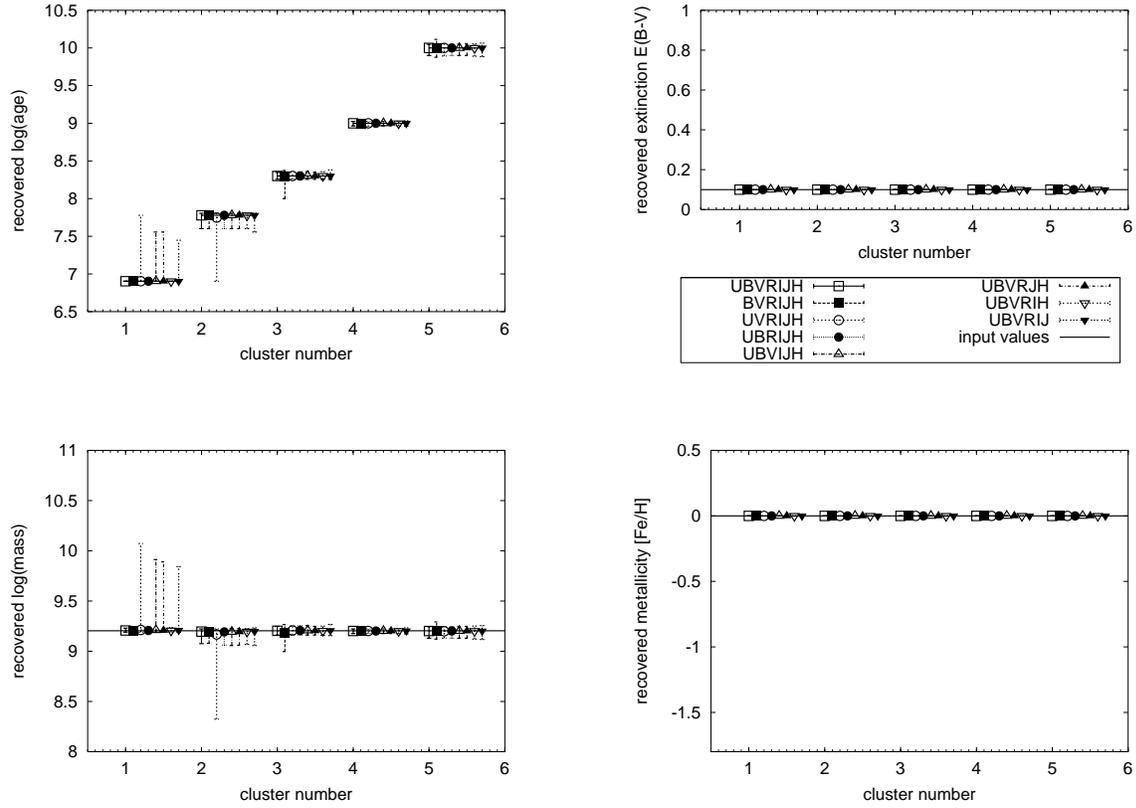}
\end{center}
\vspace{0.1cm}
\caption{Dispersion of recovered properties of artificial clusters, assuming availability of {\sl UBVRIJH} and passband combinations without one of the {\sl UBVRIJH} passbands, as indicated in the legend. Solutions were sought with extinction and metallicity fixed to the input values. Cluster parameters are standard.}
\label{fig_76ZE}
\end{figure*}

The results from the restriction of both extinction and metallicity to their
input values is presented in Fig. \ref{fig_76ZE}. Clearly, all median values
agree perfectly with the values of the remaining input parameters. The few
large uncertainty ranges indicate passband combinations that still do not
allow to determine the solutions unambiguously, because important information
(passbands) are missing. Combinations without these passbands do not allow for
a reasonable analysis. This includes combinations without the {\sl U} or {\it
B} band (and thus insufficient tracing of the kink in the SEDs) and without
the {\sl U} or {\it H} band (thereby restricting the wavelength coverage). The
{\sl R/I} bands seem to be of some importance in the early evolutionary stages
to trace the curvature of the SEDs.

Restricting the parameter space of our analysis to the input values for some
parameters clearly reproduces the input values of the others, and hence can be
used as a sanity check for the reliability of the algorithm. We find the
age-extinction degeneracy to be most important for old clusters; for such
systems a restriction in the allowed extinction range is usually possible. The
age-metallicity degeneracy is responsible for some deviations for clusters
younger than 200 Myr.

\subsection{Restricting the parameter space to incorrect values}

In this section we investigate the consequences of {\sl a priori} assuming
fixed, but incorrect generic values for the parameters.

\begin{figure*}
\begin{center}
\includegraphics[angle=-90,width=0.8\linewidth]{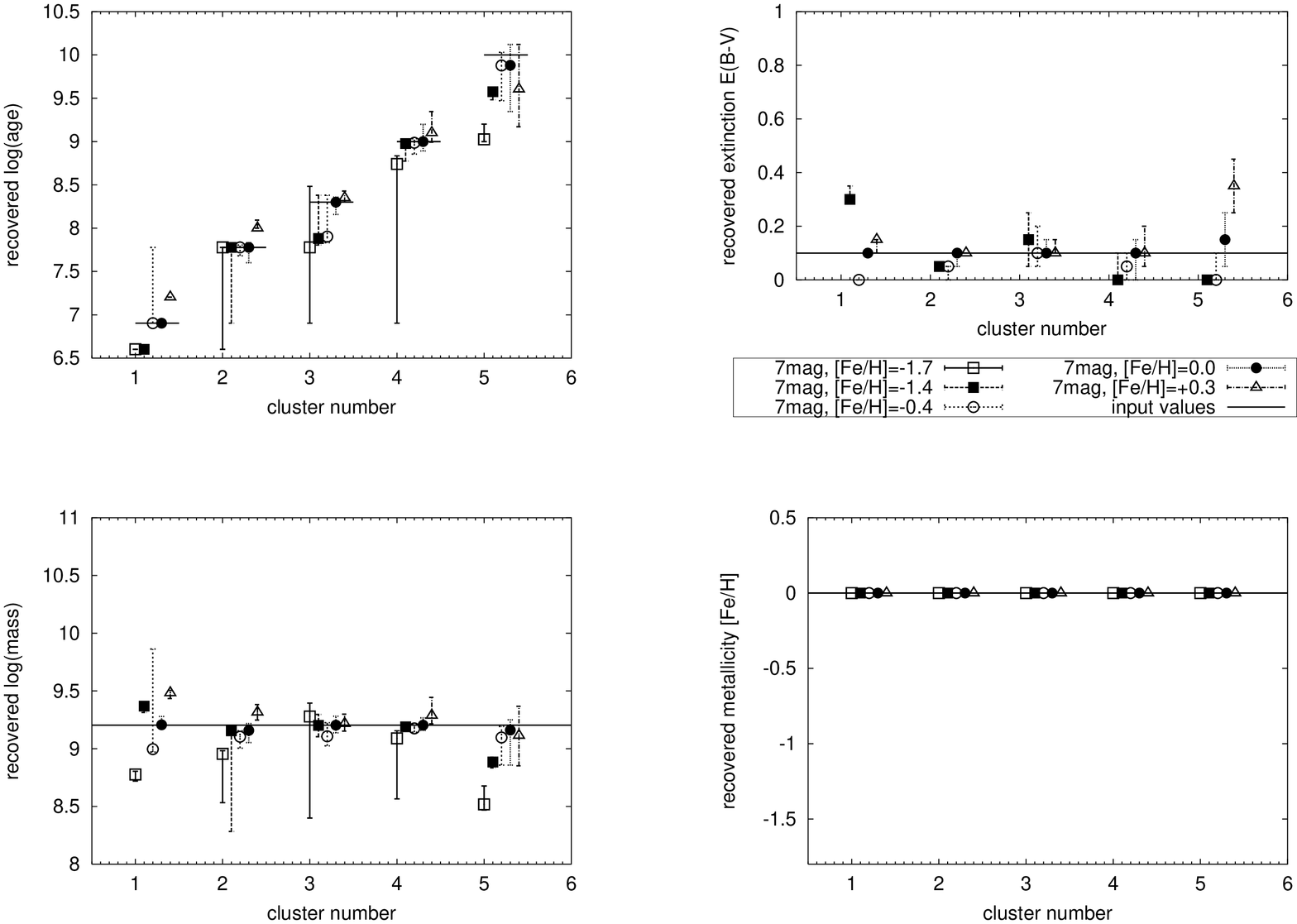}
\end{center}
\vspace{0.3cm}
\caption{Dispersion of recovered properties of artificial clusters, assuming availability of {\sl UBVRIJH} and various input metallicities, as indicated in the legend. Solutions were sought with metallicity fixed to solar metallicity. Cluster parameters are standard.}
\label{fig_wrong_7Z}
\end{figure*}

\begin{figure*}
\begin{center}
\includegraphics[angle=-90,width=0.8\linewidth]{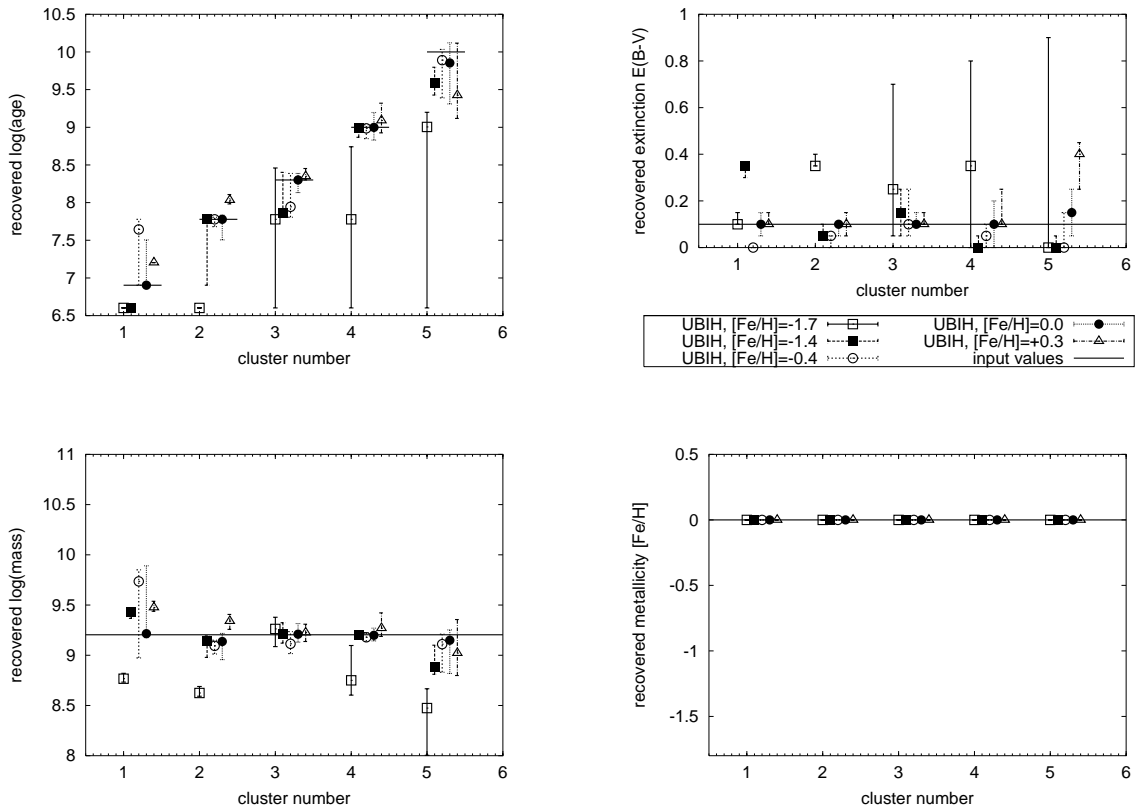}
\end{center}
\vspace{0.3cm}
\caption{Dispersion of recovered properties of artificial clusters, assuming availability of {\sl UBIH} and various input metallicities, as indicated in the legend. Solutions were sought with metallicity fixed to solar metallicity. Cluster parameters are standard.}
\label{fig_wrong_4Z}
\end{figure*}

First, we investigate the results for various input metallicities, but using
solar metallicity to recover the other input parameters (as often done in the
literature in studies of interacting and/or merging galaxies). The results are
shown in Figs. \ref{fig_wrong_7Z} and \ref{fig_wrong_4Z}. There are
significant trends in the age determination in the sense of decreasing ages
with decreasing input metallicity. These trends are in some cases accompanied
by decreasing extinction. If the actual input metallicity is lower than the
fixed metallicity assumed for the analysis, the cluster colours will be too
blue for the combined input age and extinction, and for the fixed incorrect
metallicity. Hence, either the recovered extinction is driven to lower values
and/or the solution to younger ages than their respective input values,
since both adjustments result in bluer colours for ages $\ge$ 200 Myr. The
results for an actual super-solar input metallicity can be understood the
other way around. The youngest clusters show rather randomly distributed
recovered values, thereby reflecting the complex behaviour of the magnitudes
at such young ages. Applying solar metallicity models (as often seen in the
literature) for clusters that intrinsically have sub-solar metallicity results
in ages that are too low by up to 60 per cent, masses too low by up to 56 per
cent and similarly incorrect extinction values if observations in 7 passbands
are available, and even more if the observations were obtained in only 4
passbands.

An equivalent test was done for our cluster sample in NGC 3310, where we
compared the results from assuming a fixed, solar metallicity to leaving it as
a free parameter (see de Grijs et al. 2003a). We found significantly different
age distributions in either case, with the analysis in which the metallicity
was left as a free parameter resulting in more realistic results in the
context of what is known about the starburst in NGC 3310 in general. However,
since all clusters are young in this cluster system (with ages of a few
$\times 10-100$ Myr), this test was limited to young ages.

In de Grijs et al. (2003b) we concentrated on the impact of restricting the
allowed metallicity range for the analysis. By assuming a generic, fixed
subsolar metallicity (confirmed by spectroscopy), we found that the derived
age distribution is fairly robust compared to the case where the metallicity
is left as a free parameter, but the peak of the age distribution is
significantly broadened. Hence, there is a larger dispersion for individual
clusters. This is presumably caused by clusters that do {\bf not} have the
generic metallicity value.

\begin{figure*}
\begin{center}
\includegraphics[angle=-90,width=0.8\linewidth]{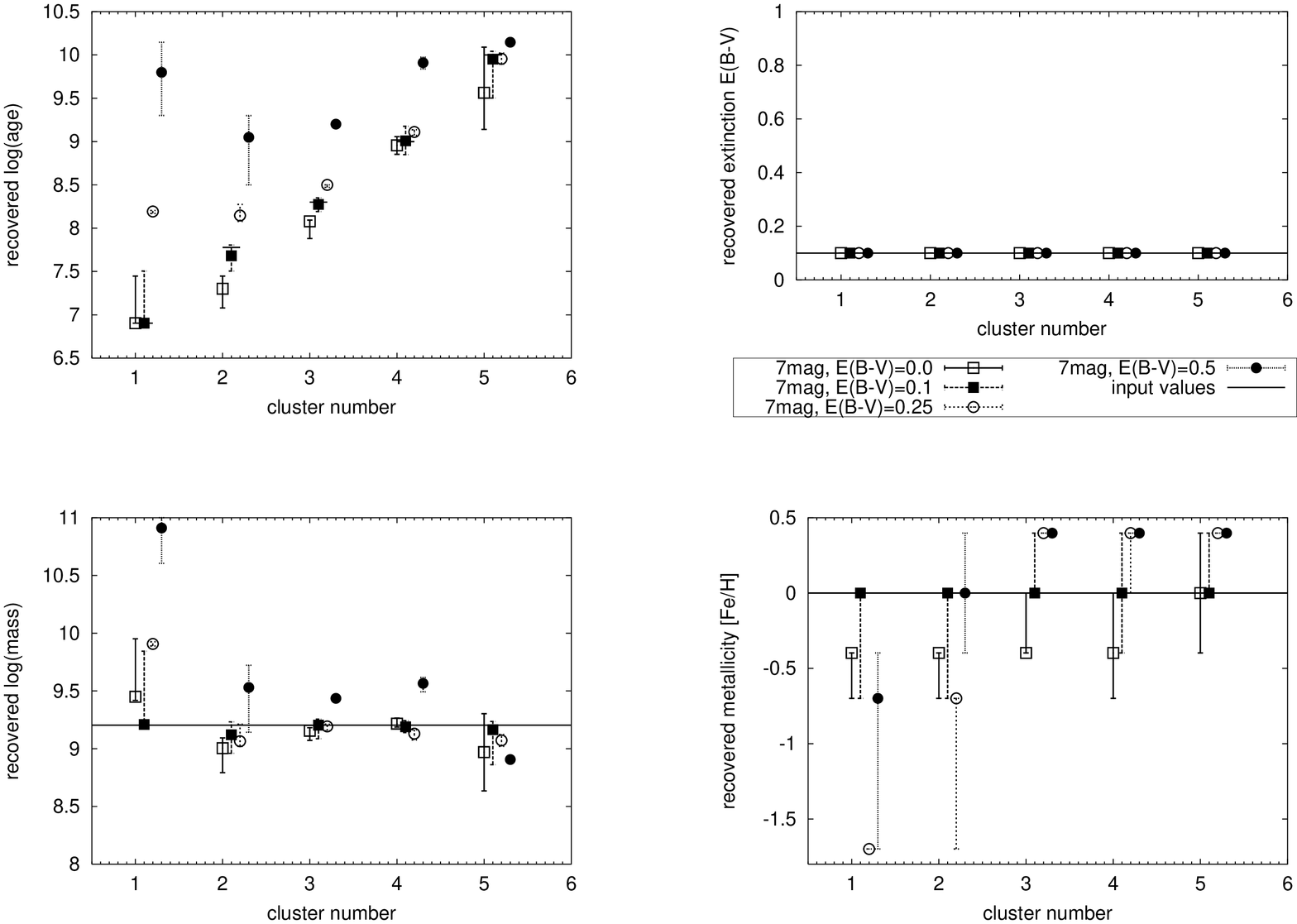}
\end{center}
\vspace{0.3cm}
\caption{Dispersion of recovered properties of artificial clusters, assuming availability of {\sl UBVRIJH} and various input extinction values, as indicated in the legend. Solutions were sought with extinction fixed to {\sl E(B-V)}=0.1. Cluster parameters are standard.}
\label{fig_wrong_7E}
\end{figure*}

\begin{figure*}
\begin{center}
\includegraphics[angle=-90,width=0.8\linewidth]{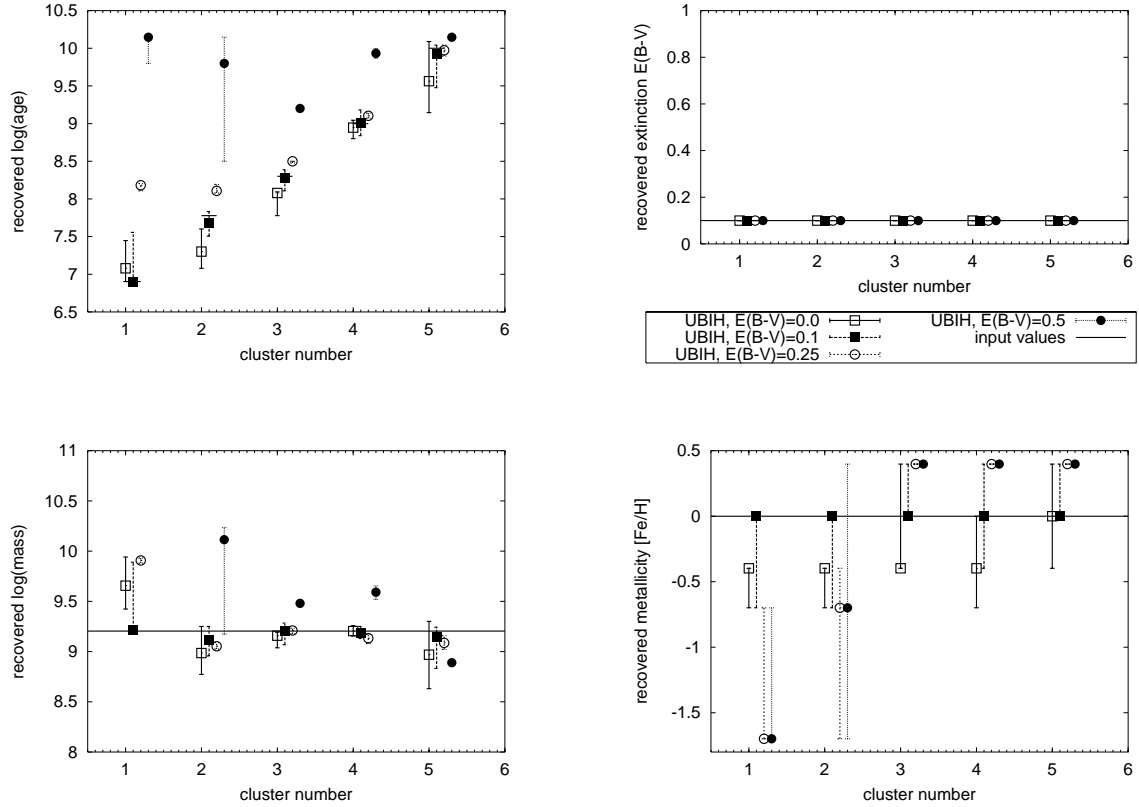}
\end{center}
\vspace{0.3cm}
\caption{Dispersion of recovered properties of artificial clusters, assuming availability of {\sl UBIH} and various input extinction values, as indicated in the legend. Solutions were sought with extinction fixed to {\sl E(B-V)}=0.1. Cluster parameters are standard.}
\label{fig_wrong_4E}
\end{figure*}

In Figs. \ref{fig_wrong_7E} and \ref{fig_wrong_4E} we show the results for
{\sl UBVRIJH} and {\sl UBIH}, respectively, if clusters with various
extinction values are analysed using a fixed value of E$(B-V)$. Shown are the
results for clusters with input values E$(B-V)=0.0, 0.1, 0.25$, and 0.5, but
analysed assuming a fixed extinction E$(B-V)_{\rm {fixed}}=0.1$. Considerable
changes are observed in the resulting metallicities and ages. In many cases,
the deviations from the input values are much larger than the derived
uncertainties. For clusters with ages $\ga$ 200 Myr there are significant
trends of increasing recovered ages and metallicities with increasing input
extinction. If the actual input extinction is smaller than the extinction
fixed for the analysis, the cluster will be too blue for the combination of
input age and metallicity, and for the fixed, incorrect extinction. Hence,
either the recovered metallicity is driven to lower values and/or the age must
be younger than the corresponding input values, since both result in bluer
colours. The results for an actual input extinction higher than the fixed
value can be understood the other way around. The youngest clusters show less
obvious trends in the distributed recovered values with increasing input
extinction, showing the complex magnitude behaviour for young ages.

{\bf The results from this section again prove the importance of determining
the physical parameters of the clusters, such as age, metallicity, and
internal extinction (and mass as a derived value), independently, and avoiding
any generic assumptions}, which might not be justified for all clusters within
a given star cluster system. This is true in particular for systems where the
existence of two distinct cluster populations is already known or suspected,
such as in merging galaxies and galaxies with known colour bimodality in their
cluster system.

\section{Conclusion}

We have presented a detailed study of the reliability and limitations of our
new algorithm to analyse observed SEDs of star clusters, based on broad-band
imaging observations, by comparing these to a grid of model SEDs from our
evolutionary synthesis code {\sc galev}.

We have computed a large grid of star cluster SEDs on the basis of our {\sc
galev} models for simple stellar populations including all relevant stellar
evolutionary phases for ages $\ge 4$ Myr. The models also include
metallicity-dependent gaseous line and continuum emission shown to be an
important contributor to broad-band fluxes in early evolutionary stages. Our
grid covers ranges in metallicity of $-1.7 \le$ [Fe/H] $\le +0.4$, in
extinction of 0 $\le$ E$(B-V) \le$ 1, and ages of 4 Myr $\le$ age $\le$ 14
Gyr. The models produce spectra from which we derive absolute magnitudes, and
hence broad-band SEDs, for any given filter system. Here, we present results
for {\sl HST} broad-band filters widely used for observations and analyses of
star cluster systems in external galaxies.

Our parameter analysis algorithm compares a given cluster SED (either observed 
or theoretical, as done in this study)
with the model SEDs from our input parameter grid. Each parameter set is
assigned a certain probability, based on an ``observation--model'' comparison
using a chi--squared algorithm. The parameter set with the highest probability is
adopted as the best model; the range of parameters from sets with the highest
probabilities (up to a total probability of 68.26 per cent) determines the
$1\sigma$ uncertainties in the parameters.

We constructed numerous artificial cluster SEDs, and varied each of the input
parameters in turn to assess their effects on the robustness of our parameter
recovery. For each clean model artificial cluster SED we calculated 10,000
additional clusters, with errors distributed around the input magnitudes in a
Gaussian fashion.

We identified useful and less suitable passband combinations, with the aim to
aid in the planning of observational campaigns. Although a large number of
passbands is always preferable, any realistic programme will more likely be
limited to observations in the minimum number of required passbands to
successfully reach its goals. {\bf In order to successfully disentangle the
three free parameters age, metallicity, and internal extinction based on the
shape of a broad-band SED, and to determine the mass of a star cluster by
simple scaling of the model magnitudes to the observed level, a minimum of
four passbands are required.} The most/least preferable passband combinations
are summarised in Tables 1 and 2 as a function of the expected age of the
cluster population. {\bf In all cases, tracing the kink (or hook) in the SEDs
around the {\sl B} band (see Fig. \ref{fig_SED}) is of the highest importance.
The inclusion of at least one NIR passband significantly improves the
results}, since NIR wavelengths allow to efficiently restrict the metallicity
range. For the youngest clusters, metallicity estimates are determined by the
{\sl U} and/or {\it B} bands. The poorest results are obtained if neither UV
information nor {\sl B} band data are available, or if the available
wavelength coverage is very short or biased towards blue or red
wavelengths (like {\sl RIJH}).

\begin{table}
\label{tab1}
\begin{center}
\caption{Overview of the most important filters and most/least preferable 4-passband combinations, if NIR data are available}
\begin{tabular}{|c|c|c|c|}
\hline
 & & &\\ 
 Age & important & preferable & combinations\\
 & filters & combinations & to be avoided\\
 & & &\\ 
\hline
 & & &\\
 $\le$ few Gyr& {\sl U, B}    & {\sl UBIH, UBVH} & {\sl BVIH, RIJH}\\
 $\ge$ few Gyr& {\sl B, V, U} & {\sl BVIH, UBVI} & {\sl UVIH, UBIH}\\
 & & &\\
\hline
\end{tabular}
\end{center}
\end{table}

\begin{table}
\label{tab2}
\begin{center}
\caption{Overview of the most important filters and most/least preferable 4-passband combinations, if no NIR data are available}
\begin{tabular}{|c|c|c|c|}
\hline
 & & &\\ 
 Age & important & preferable & combinations\\
 & filters & combinations & to be avoided\\
 & & &\\ 
\hline
 & & &\\
 $\le$ few Gyr& {\sl U, B}    & {\sl UBRI, UBVI} & {\sl BVRI, UVRI}\\
 $\ge$ few Gyr& {\sl B, V, U} & {\sl UBVI}       & {\sl UVRI, UBRI}\\
 & & &\\
\hline
\end{tabular}
\end{center}
\end{table}

By analysing artificial clusters, using a variety of input parameters
(specifically age, metallicity, and internal extinction) with our new code, we
find in general good agreement between the recovered and the input parameters.
Only the oldest, 10 Gyr-old artificial clusters show significant signs of the
well-known age-metallicity-extinction degeneracy.

We have considered several {\sl a priori} restrictions of the parameter space,
both to the (correct) input values and to some commonly assumed values. We
easily recover all remaining input values correctly if one of them is
restricted, {\sl a priori} to its correct input value; this also provides a
sanity check for the reliability of our code. We find the age-extinction
degeneracy to be most important for old clusters. For such systems, an {\sl a
priori} restriction of the allowed extinction range is often possible and
shown to be very useful. The age-metallicity degeneracy is responsible for
some misinterpretations of clusters younger than 200 Myr.

If we, however, restrict one or more of our input parameters {\sl a priori} to
incorrect values (such as using, e.g., only solar metallicity, as often found
in the literature), large uncertainties result in the remaining parameters.
While certain restrictions might be justified in specific cases, we strongly
advice caution in more complex cases, such as in interacting galaxies or in
galaxies with known colour bimodality in their cluster systems.

{\bf Finally, we conclude that reliable determination of physical star cluster
parameters is possible on the basis of broad-band imaging, provided the
availability of a useful set of observational passbands, containing at least
four filters, a sufficiently long wavelength base line, and reasonable
photometric accuracy. We show that a small, but suitably chosen filter set
with deep observations (and the correspondingly small uncertainties) gives
more reliable results than a larger number of shallow exposures in
inappropriate or redundant filters.}

The method we have developed is a versatile and useful tool for the
interpretation of large multi-colour data sets for star clusters of different
ages and in a large variety of environments, such as provided by, e.g., our
ST-ECF/ESO {\sc astrovirtel}\footnote{ASTROVIRTEL is a project funded by the European 
Commission under 5FP contract HPRI-CT-1999-00081.} project 
``The Evolution and Environmental Dependence of
Star Cluster Luminosity Functions'' (PI R. de Grijs).


\section{Acknowledgements}
PA is partially supported by DFG grant Fr 916/11-1.

\end{document}